\documentclass[a4paper,11pt]{article}

\usepackage{jheppub}
\usepackage{graphicx}
\usepackage{amsmath}
\usepackage{slashed}
\usepackage{braket}
\usepackage[export]{adjustbox}
\usepackage{enumitem}
\usepackage{placeins}
\usepackage[normalem]{ulem}
\usepackage{xspace}

\newcommand{\refcite}[1]{ref.~\cite{#1}}
\newcommand{\refscite}[1]{refs.~\cite{#1}}
\newcommand{\Eq}[1]{Eq.~\eqref{eq:#1}}
\newcommand{\eq}[1]{eq.~\eqref{eq:#1}}
\newcommand{\eqs}[2]{eqs.~\eqref{eq:#1} and \eqref{eq:#2}}
\renewcommand{\sec}[1]{section~\ref{sec:#1}}

\newcommand{\app}[1]{appendix~\ref{app:#1}}
\newcommand{\fig}[1]{figure~\ref{fig:#1}}

\newcommand{\tbl}[1]{table~\ref{tbl:#1}}

\newcommand{\df}{\mathrm{d}}
\newcommand{\img}{\mathrm{i}}

\newcommand{\zb}{\bar z}

\newcommand{\bn}{{\bar n}}
\newcommand{\bq}{{\bar q}}

\newcommand{\GeV}{\,\mathrm{GeV}}

\newcommand{\cJ}{\mathcal{J}}
\newcommand{\cK}{\mathcal{K}}
\newcommand{\cL}{\mathcal{L}}

\newcommand{\cN}{\mathcal{N}}
\newcommand{\cO}{\mathcal{O}}

\newcommand{\qt}{{\vec q}_T}
\newcommand{\bt}{{\vec b}_T}

\newcommand{\as}{\alpha_s}
\newcommand{\GammaC}{\Gamma_{\rm cusp}}
\newcommand{\nn}{\nonumber}

\newcommand{\lqcd}{\Lambda_\mathrm{QCD}}

\newcommand{\SYM}{$\mathcal{N}=4$~sYM\xspace}

\newcommand{\moment}{$N=2$ Mellin moment }
\newcommand{\born}{ \hat \sigma_0}

\newcommand{\dsz}{\frac{\df \sigma_0}{\df z}}
\newcommand{\dso}{\frac{\df \sigma_1}{\df z}}
\newcommand{\dsreg}{\frac{\df \sigma_{\text{reg.}}}{\df z}}

\def\beq{\begin{equation}}
\def\eeq{\end{equation}}
\def\bea{\begin{eqnarray}}
\def\eea{\end{eqnarray}}

\newcommand{\EEC}{\mathrm{EEC}}
\newcommand{\EECreg}{\frac{\df \sigma^\text{reg}}{\df z}}
\newcommand{\EECregas}[1]{\left.{\EECreg}\right|_{\cO(\alpha_s^{#1})}}
\newcommand{\EECregaslab}[2]{\left.{\frac{\df \sigma_{#2}^\text{reg}}{\df z}}\right|_{\cO(\alpha_s^{#1})}}

\title{\boldmath The Energy-Energy Correlation in the back-to-back limit at N$^3$LO and N$^3$LL$^\prime$}

\author[a,b]{Markus A.~Ebert,}
\emailAdd{ebert@mpp.mpg.de}

\author[c]{Bernhard Mistlberger,}
\emailAdd{bernhard.mistlberger@gmail.com}

\author[c]{and Gherardo Vita}
\emailAdd{gherardo@slac.stanford.edu}

\affiliation[a]{Max-Planck-Institut f\"ur Physik, F\"ohringer Ring 6, 80805 M\"unchen, Germany}
\affiliation[b]{Center for Theoretical Physics, Massachusetts Institute of Technology, Cambridge, Massachusetts 02139, USA}
\affiliation[c]{SLAC National Accelerator Laboratory, Stanford University, Stanford, CA 94039, USA}

\abstract{
We present the analytic formula for the Energy-Energy Correlation (EEC) in electron-positron annihilation computed in perturbative QCD to next-to-next-to-next-to-leading order (N$^3$LO) in the back-to-back limit.
In particular, we consider the EEC arising from the annihilation of an electron-positron pair into a virtual photon as well as a Higgs boson and their subsequent inclusive decay into hadrons.
Our computation is based on a factorization theorem of the EEC formulated within Soft-Collinear Effective Theory (SCET) for the back-to-back limit.
We obtain the last missing ingredient for our computation - the jet function - from a recent calculation of the transverse-momentum dependent fragmentation function (TMDFF) at N$^3$LO.
We combine the newly obtained N$^3$LO jet function with the well known hard and soft function to predict the EEC in the back-to-back limit.
The leading transcendental contribution of our analytic formula agrees with previously obtained results in $\cN = 4$ supersymmetric Yang-Mills theory.
We obtain the \moment of the bulk region of the EEC using momentum sum rules.
Finally, we obtain the first resummation of the EEC in the back-to-back limit at N$^3$LL$^\prime$ accuracy, resulting in a factor of $\sim 4$ reduction of uncertainties in the peak region compared to N$^3$LL predictions.
}

\preprint{\vbox{%
\hbox{MIT--CTP/5263}
\hbox{SLAC--PUB--17579}
\hbox{MPP--2020--225}
}}

\begin{document}

\maketitle
\newpage

\section{Introduction}

In the quest of understanding the nature of the strong interaction, the study of QCD radiation produced in high-energy electron-positron collisions provides a powerful lens into the behavior of Quantum Chromodynamics (QCD).
Among the observables that one can consider to perform precision studies of QCD radiation,
the Energy-Energy Correlation (EEC) \cite{Basham:1978bw} stands out for its simplicity,
and has been an important benchmark observable to test QCD and to extract the strong coupling constant $\as$,
which has been explored both at LEP and at SLAC~\cite{Abreu:1990us,Acton:1991cu,Acton:1993zh,Abreu:1993kj,Abe:1994mf,Tulipant:2017ybb,Kardos:2018kqj,dEnterria:2019its}.

The EEC is an $e^+ e^-$ event shape to measure the energy-weighted angular distance between any pair of particles in an event~\cite{Basham:1978bw}, and as such is one of the earliest examples of an infrared and collinear (IRC) safe observable. It is defined as
\begin{align}
 \EEC(\chi) =  \frac{\df \sigma}{\df \chi}= \sum_{i,j} \int \df \sigma_{e^+e^-\to i j + X} \, \frac{E_i E_j}{Q^2} \, \delta(\cos\theta_{ij} - \cos \chi)
\,.\end{align}
Here, the sum runs over all pairs of particles $\{i,j\}$ in the final state, with $E_{i,j}$ denoting their energies,
$Q^2$ is the invariant mass of the $e^+ e^-$ collision, and $\theta_{ij}$ is the angle between the particles.
The differential cross section $\df\sigma_{e^+e^-\to i j + X}$ contains the phase-space measure and squared matrix elements for the process $e^+ e^- \to ij + X$.

An extensive effort has been devoted to the calculation of the EEC, both in QCD and in maximally supersymmetric Yang-Mills theory (\SYM).
In QCD, results for the EEC were obtained numerically in \refcite{DelDuca:2016csb,Tulipant:2017ybb} at NNLO in the CoLoRFulNNLO framework~\cite{Somogyi:2006da,Somogyi:2006db,Aglietti:2008fe}. The analytic form of the EEC is only known at NLO in QCD, thanks to the recent calculation of~\refcite{Dixon:2018qgp}, see also \refcite{Luo:2019nig} for the EEC in gluon-initiated Higgs decays (sometimes referred to as the Higgs EEC).
In \SYM, the EEC was calculated analytically both at NLO~\cite{Belitsky:2013xxa,Belitsky:2013bja,Belitsky:2013ofa}
and at NNLO~\cite{Henn:2019gkr},
and also at strong coupling using the AdS/CFT correspondence~\cite{Maldacena:1997re,Hofman:2008ar}.
Moreover, much progress has been achieved in understanding the EEC in $\cN=4$ SYM \cite{Hofman:2008ar,Belitsky:2013xxa,Belitsky:2013bja,Belitsky:2013ofa,Henn:2019gkr,Moult:2019vou} and interesting efforts are being employed to shed light on its relation to energy correlators in QCD~\cite{Korchemsky:2019nzm,Dixon:2019uzg,Chicherin:2020azt,Henn:2020omi,Chen:2020uvt,Chen:2020adz}.
Another observable closely related to the EEC is the Transverse Energy Energy Correlator (TEEC)~\cite{Ali:1984yp}. A factorization theorem for the TEEC in the back-to-back limit has been presented in \refcite{Gao:2019ojf} for hadron-hadron colliders and extended in \refcite{Li:2020bub} for DIS,%
\footnote{For a recent analysis of the TEEC at the electron-proton collider HERA see \refcite{Ali:2020ksn}.}
which shares various ingredients with the factorization of the EEC.

The EEC is commonly expressed in the variable $z$,
\begin{align} \label{eq:z}
 z \equiv \frac12 (1 - \cos\chi)
\,.\end{align}
The differential cross section is distribution-valued in the small-angle limit (sometimes also referred to as collinear or forward limit) $\chi\to0$ and in the back-to-back limit $\chi\to\pi$.
By expressing the EEC in terms of $z$ the singular points of the distributions are mapped onto $z\to0$ and $z\to1$, respectively.
The differential cross section at Born level is given by
\beq
\frac{\df \sigma}{\df z}=\frac{2}{\sin \chi}\frac{\df \sigma}{\df\chi}=\frac{1}{2 } \born \delta(z)+\frac{1}{2 }\delta(1-z)\born+\mathcal{O}(\as),
\eeq
where $\as$ is the strong coupling constant.
In this article we will direct attention towards the singular limits of the EEC. 
Consequently, it is useful to split the cross section into three different contributions.
\begin{align}
 \frac{\df \sigma}{\df z} =\dsz+\dsreg+\dso
\,,\end{align}
where $\dsz$ and $\dso$ contain all terms that behave as $1/z$ and $1/(1-z)$, respectively.
More precisely, they are expressed in terms of (plus) distributions to regulate these divergences,
while $\dsreg$ is a regular function of $z$ that is holomorphic in the entire unit interval, $z\in[0,1]$.

The EEC can also be expressed in the variable 
\beq
\frac{\df \sigma}{\df x}=2 x \frac{\df \sigma}{\df z},\hspace{1cm}x=\sqrt{z}=\sin\left(\frac{\chi}{2}\right),\hspace{1cm}x\in[0,1].
\eeq
The gluon and photon-induced EEC was computed through order $\as^2$ in refs.~\cite{Dixon:2018qgp,Luo:2019nig} in terms of classical polylogarithms as a function of $z$.
Here, we note that expressing the differential cross section in terms of the variable $x$ allows us to represent it in terms of harmonic polylogarithms~\cite{Remiddi:1999ew} with argument $x$ and indices $\{-1,0,1\}$.
We provide analytic formulae for the EEC expressed in $x$ including all distribution valued terms through $\as$ as ancillary files together with the arXiv submission of this article.
The recent computation of the EEC at $\mathcal{O}(\as^3)$ in \SYM in ref.~\cite{Henn:2019gkr} finds evidence that at this order the analytic formula contains elliptic functions and is no longer expressible in terms of HPLs.

At Born level the EEC vanishes for $z \ne 0,1$.
Consequently, in the context of the fixed-order calculations reported above it is customary not to include the distributional behavior as $z\to0,1$ and count $\cO(\as)$ as the leading order (LO).
Accordingly $\cO(\as^{n+1})$ contributions are counted as N$^n$LO.
Such fixed-order calculations become unreliable in the singular limits $z\to0$ and $z\to1$, where large logarithms $\ln(z)$ and $\ln(1-z)$ can spoil perturbative convergence.
In these limits the EEC needs to be resummed to all orders in perturbation theory to retain predictive power.

In the forward limit, this resummation was performed at leading logarithmic (LL) accuracy a long time ago~\cite{Konishi:1978ax}.
Recently, a factorization theorem was derived in this limit and the resummation was improved in ref.~\cite{Dixon:2019uzg}.
In the back-to-back limit, the resummation was carried out at next-to-next-to-leading logarithmic (NNLL) accuracy~\cite{deFlorian:2004mp,Tulipant:2017ybb}
based on transverse-momentum dependent (TMD) factorization in $e^+ e^-$~\cite{Collins:1981uk,Collins:1981va,Kodaira:1981nh,Kodaira:1982az}.
Recently, \refcite{Moult:2018jzp} proofed that the all-order factorization for the EEC in the back-to-back limit indeed follows from TMD factorization, and presented first results at N$^3$LL acurracy.
In \SYM, the factorization of the EEC in both its forward and back-to-back limit has also been explored up to four loops by using the operator product expansion for light-ray operators~\cite{Kologlu:2019mfz}, and by relating the EEC to four-point correlation functions of conserved currents~\cite{Korchemsky:2019nzm}.
Note that these factorization theorems contain the full distributional structures,
and thus their fixed-order expansions start at $\cO(\as^0)$.
Hence, in the context of factorization theorems one counts $\cO(\as^n)$ as N$^n$LO accuracy.

In this paper we calculate the full singular structure of the EEC in QCD in the back-to-back limit at N$^3$LO, i.e. $\cO(\alpha_s^3)$.
This is achieved by calculating the EEC jet function at the same order, which is the only unknown ingredient of the factorization theorem derived in \refcite{Moult:2018jzp}.
As an application, we resum the EEC in the back-to-back limit at N$^3$LL$^\prime$ accuracy,
i.e.~N$^3$LL resummation combined with the full N$^3$LO fixed-order boundary condition.
This constitutes the highest level of accuracy obtained for an event shape sensitive to QCD radiation to date.
We show that thanks to the resummation of large logarithms up to N$^3$LL and the inclusion of fixed-order boundary terms at N$^3$LO, we achieve a $\sim 4$-fold reduction of uncertainty compared to previous results obtained at lower accuracy.
We thoroughly discuss different schemes to estimate the uncertainties due to missing higher order corrections both in the boundary terms as well as in the anomalous dimensions and compare them with the ones used in the literature for this observable. Adopting a scheme in line with the ones previously used in the literature we obtain a $0.5\%$ uncertainty at the peak.
Using a more conservative scheme, which also estimates uncertainties from soft physics, we obtain a 4\% uncertainty for our result at N$^3$LL$^\prime$.
Furthermore, using a momentum sum rule we also analytically calculate the \moment of the regular part at N$^3$LO,
which will be an important check of the full result once it becomes available.
The jet functions calculated in this work are necessary ingredients to describe the singular behavior of the TEEC in the large angle limit at $\cO(\alpha_s^3)$ as well as to the resummation of large logs at N$^3$LL$^\prime$ and N$^4$LL accuracy for this observable.

This paper is organized as follows.
In \sec{EEC} we briefly review the factorization theorem for the EEC in the back-to-back
and show how to extend it to gluon-induced Higgs decays, clarifying its nontrivial helicity structure.
We also obtain the three-loop jet functions from our results for the transverse-momentum dependent fragmentation functions (TMDFFs) calculated in the companion paper~\cite{Ebert:2020qef},%
\footnote{While \refcite{Ebert:2020qef} was finalized, an independent calculation of the TMDFF at N$^3$LO also appeared in \refcite{Luo:2020epw}.}
and present the full singular structure of the EEC at N$^3$LO in QCD in the back-to-back limit.
In addition, we validate the conjecture that in the back-to-back limit, the leading transcendental terms in QCD
match those in $\cN=4$ supersymmetric Yang-Mills \refcite{Korchemsky:2019nzm}.
In \sec{sum_rules}, we exploit the fact that the EEC obeys a set of sum rules to analytically obtain the \moment of the bulk of the EEC distribution at NNLO in QCD.
In \sec{resummation}, we carry out the resummation of the EEC at N$^3$LL$^\prime$ accuracy to illustrate the improved perturbative accuracy compared to previous results.
We conclude in \sec{conclusion}.

\section{The EEC in the back-to-back limit}
\label{sec:EEC}

\subsection{Factorization of the EEC in the back-to-back limit}\label{sec:fact_intro}

In this section, we briefly review the factorization of the EEC in the back-to-back limit $z\to1$.
In the case of electron-positron annihilation, i.e.~quark-initiated EEC, this was first derived by Collins and Soper~\cite{Collins:1981uk, Collins:1981va} and Kodaira and Trentadue~\cite{Kodaira:1981nh,Kodaira:1982az}.
Recently, it was also formulated in \refcite{Moult:2018jzp} using Soft-Collinear Effective Theory (SCET) \cite{Bauer:2000ew, Bauer:2000yr, Bauer:2001ct, Bauer:2001yt},
which clarified the role of nontrivial fixed-order boundary terms in the factorization formula.
So far, no details have been given in the literature on the Higgs EEC.
To fill this gap, we briefly review the derivation of the quark EEC in \sec{fact_EEC_q},
and then show how the same strategy can be applied to the gluon EEC in \sec{fact_EEC_g},
where additional complications arise from a nontrivial Lorentz structure.

\subsubsection{Quark EEC}
\label{sec:fact_EEC_q}

In the back-to-back limit, the EEC can be related to identified hadron production,
$e^+ e^- \to h_1 h_2 + X$, at small transverse momentum $\qt$ of the dihadron system~\cite{Collins:1981uk, Collins:1981va, Moult:2018jzp},
\begin{align} \label{eq:EEC_qT_fact}
 \lim_{z\to1} \frac{\df\sigma}{\df z}
 = \int_0^1 \df z_1 \df z_2 \, \frac{z_1 z_2}{2} \int\df^2\qt \, \delta\biggl(1 - z - \frac{q_T^2}{Q^2} \biggr)
   \lim_{q_T\to0} \sum_{h_1, h_2} \frac{\df\sigma_{e^+ e^- \to h_1 h_2}}{\df z_1 \df z_2 \df^2\qt}
\,.\end{align}
Here, $z_{1,2} = (P_{1,2} \cdot q)/ q^2$ describe the longitudinal momenta carried away by the hadrons,
with the total momentum transfer given by $q^\mu = p_{e^+}^\mu + p_{e^-}^\mu$.%
\footnote{In a frame where the hadrons are aligned along back-to-back lightlike directions,
i.e.~$p_1^\mu = p_1^- n^\mu/2$ and $p_2^\mu = p_2^+ \bn^\mu/2$ with $n^2 = \bn^2 = 0$ and $n \cdot \bn = 2$,
these evaluate to $z_1 = p_1^-/q^-$ and $z_2 = p_2^+/q^+$, up to corrections suppressed by $q_T$.
In this frame, $\qt$ is the transverse component of the momentum transfer $q^\mu$.}
The factorization of dihadron production at small $q_T$ was derived in seminal works by Collins and Soper~\cite{Collins:1981uk, Collins:1981va}
(see also \refcite{Collins:2011zzd}), and can be written as
\begin{align} \label{eq:qT_fact_q}
 \frac{\df\sigma_{e^+ e^- \to h_1 h_2}}{\df z_1 \df z_2 \df^2\qt} &
 = \born H_{q\bar q}(Q, \mu)
   \int\!\frac{\df^2\bt}{(2\pi)^2} \, e^{\img \qt \cdot \bt}
   \tilde D_{h_1 / q}\Bigl(z_1, b_T, \mu, \frac{\nu}{Q}\Bigr) \tilde D_{h_2 / \bar q} \Bigl(z_2, {b_T}, \mu, \frac{\nu}{Q}\Bigr)
   \nn\\&\quad \times
   \tilde S_q(b_T, \mu, \nu)
   ~\times~\Bigl[1 + \cO\Bigl(\frac{q_T^2}{Q^2}\Bigr)\Bigr]
\,.\end{align}
Here, summation over all quark flavors $q$ is kept implicit, the hard function $H_{q\bar q}$ encodes virtual corrections to Born process $e^+ e^- \to q \bar q$, $\tilde D_{h / q}$ is the fragmentation function encoding the probability to obtain the hadron $h$ from the fragmentation of a quark $q$, and $\tilde S_q$ is the TMD soft function.%
\footnote{The TMD soft function is universal between $e^+ e^-$ and $pp$ processes~\cite{Collins:2004nx},
i.e.\ it is independent of the direction of the contained Wilson lines, and thus we do not further distinguish these.
For a detailed discussion of the equivalence of the TMD soft function in the context of the EEC, see \refcite{Moult:2018jzp,Zhu:2020ftr}.}
Note, that in the literature, the soft function is often combined with the fragmentation function as $\tilde D_{h / q} \sqrt{\tilde S_q}$, whereas we keep it explicit.
The scale $\nu$ is the so-called rapidity renormalization scale, which is closely related to the Collins-Soper scale $\zeta$.
Note that the fragmentation functions $\tilde D_{h/q}$ and the soft function $\tilde S_q$ both depend on the chosen rapidity renormalization scheme.
This scheme dependence cancels in the combination $\tilde D_{h / q} \sqrt{\tilde S_q}$,
and consequently also in the combination $\tilde D_{h_1/q} \tilde D_{h_2/\bq} \tilde S_q$ in \eq{qT_fact_q}.
For more details on the fragmentation function $\tilde D_{h/q}$, we refer to \refcite{Ebert:2020qef}.

By combining \eqs{EEC_qT_fact}{qT_fact_q}, we obtain the EEC factorization theorem in the back-to-back limit as stated in \refcite{Moult:2018jzp},
\begin{align} \label{eq:EEC_fact_thm_q}
 \frac{\df\sigma}{\df z} &
 = \frac{\born}{2} H_{q\bar q}(Q, \mu)
   \int\frac{\df^2\bt \, \df^2\qt}{(2\pi)^2} e^{\img \qt \cdot \bt} \delta\biggl(1 - z - \frac{q_T^2}{Q^2} \biggr)
   J_q\Bigl(b_T, \mu, \frac{\nu}{Q}\Bigr) J_\bq\Bigl(b_T, \mu, \frac{\nu}{Q}\Bigr) \tilde S_q(b_T, \mu, \nu)
   \nn\\&\quad \times [ 1 + \cO(1-z) ]
\nn\\&
 = \frac{\born}{8} H_{q\bar q}(Q,\mu) \int_0^\infty \df (b_T Q)^2 \, J_0\bigl(b_T Q \sqrt{1-z}\bigr)
   J_q\Bigl(b_T, \mu, \frac{\nu}{Q}\Bigr) J_\bq\Bigl(b_T, \mu, \frac{\nu}{Q}\Bigr) \tilde S_q(b_T, \mu, \nu)
   \nn\\&\quad \times [ 1 + \cO(1-z) ]
\,,\end{align}
where $H_{q\bq}$ and $\tilde S_q$ are the same hard and soft functions as in \eq{qT_fact_q},
$J_0(x)$ is the $0$-th Bessel function of the first kind, and the jet functions $J_q$ are defined as
the first moments of the fragmentation functions,
\begin{align} \label{eq:def_Jq}
 J_q\Bigl(b_T, \mu, \frac{\nu}{Q}\Bigr) &\equiv
 \sum_{h} \int_0^1 \df z \, z  \, \tilde D_{h / q}\Bigl(z, b_T, \mu, \frac{\nu}{Q}\Bigr)
\,.\end{align}
The jet function $J_q$ has the same dependence on the rapidity renormalization scheme as the fragmentation function $\tilde D_{h/q}$.
Similar to \eq{qT_fact_q}, this scheme dependence cancels in the combination $J_q J_\bq \tilde S_q$ that arises in \eq{EEC_fact_thm_q}.
In analogy to TMD factorization, where it is often customary to absorb the soft function in the TMDPDF or TMDFF,
one can construct a manifestly rapidity-regulator independent jet function as
\begin{align} \label{eq:calJdef}
 \cJ_q(b_T,\mu,\zeta) = J_q\left(b_T,\mu,\frac{\nu}{Q}\right) \sqrt{\tilde S_q\left(b_T,\mu,\nu\right)}
\,,\end{align}
where the Collins-Soper scale $\zeta^2 = Q$ as in TMD factorization.
Using \eq{calJdef}, \eq{EEC_fact_thm_q} can be equivalently written as
\begin{align} \label{eq:EEC_fact_thm_q_calJ}
 \frac{\df\sigma}{\df z} &
 = \frac{\born}{8} H_{q\bar q}(Q,\mu) \int_0^\infty \df (b_T Q)^2 \, J_0\bigl(b_T Q \sqrt{1-z}\bigr)
   \cJ_q(b_T,\mu,\zeta) \cJ_\bq(b_T,\mu,\zeta) \times [ 1 + \cO(1-z) ]
\,,\end{align}
For more details on this construction in the context of TMDPDFs and TMDFFs, see \refscite{Ebert:2020qef, Ebert:2020yqt}.
In the following, we will restrict our discussion to $J_q$, as the corresponding results for $\cJ_q$ can be trivially obtained using \eq{calJdef}.

To simplify the jet functions, we use that for perturbative $b_T \lesssim \lqcd^{-1}$ they can be perturbatively
matched onto the collinear fragmentation functions $d_{h / q}$ as
\begin{align} \label{eq:F_matching}
 \tilde D_{h / q}\Bigl(z, b_T, \mu, \frac{\nu}{Q}\Bigr)
 = \sum_{q'} \int_z^1 \frac{\df z'}{z'} d_{h/q'}\Bigl(\frac{z}{z'}\Bigr)
   \tilde \cK_{q q'}\Bigl(z', b_T, \mu, \frac{\nu}{Q}\Bigr)
   + \cO(b_T^2 \lqcd^2)
\,,\end{align}
where $\tilde \cK_{q q'}$ is a perturbative matching kernel.
Applying \eq{F_matching} to \eq{def_Jq}, we obtain
\begin{align} \label{eq:TMDFFtoJ}
 J_q\Bigl(b_T, \mu, \frac{\nu}{Q}\Bigr) &
  = \sum_{q'} \int_0^1 \df z' \, z' \, \tilde \cK_{q q'}\Bigl(z', b_T, \mu, \frac{\nu}{Q}\Bigr)
\,,\end{align}
where we used the momentum sum rule of the fragmentation function,
\begin{align} \label{eq:sum_rule}
 \sum_h \int_0^1 \df z \, z \, d_{h / q}(z,\mu) = 1
\,.\end{align}
Thus, the EEC jet function is free from nonperturbative hadronic matrix elements,
which makes the EEC much less susceptible to nonperturbative effects than the $q_T$ distribution itself.
We note that perturbative power corrections to \eq{EEC_fact_thm_q} can be systematically studied using the operator formalism of SCET~\cite{Moult:2019vou,Feige:2017zci,Moult:2017rpl} and involve the treatment of rapidity divergences beyond leading power~\cite{Ebert:2018gsn,Moult:2017xpp}.
Nonperturbative power corrections to the EEC have been explored in \refscite{Korchemsky:1999kt,Li:2021txc}.

The factorization for the EEC in the back-to-back limit has been explored in the literature since a long time.
In \refscite{Collins:1981uk,Collins:1981va}, Collins and Soper derived the $q_T$ factorization theorem in \eq{qT_fact_q},
which albeit using a different notation already contained hard and transverse-momentum dependent fragmentation functions,
with the soft function absorbed into the TMDFFs. They also showed how to resum large logarithms by solving evolution equations
in the unphysical scales $\mu$ and $\nu$, which we will discuss in detail in \sec{resummation}.
Using the relation between $z$ and small $q_T$, they also obtained a resummed formula for the EEC.
However, it was only provided at LO in the matching, where the hard, jet and soft functions all evaluate to unity.
Similarly, Kodaira and Trentadue presented $q_T$ factorization with TMDFFs that are matched onto collinear FFs,
but only provided formulas for the resummed EEC spectrum valid at LL and NLL,
where the hard and jet functions again evaluate to unity~\cite{Kodaira:1981nh,Kodaira:1982az}.
Nontrivial fixed-order terms in the factorization formula were first pointed out in \refcite{deFlorian:2004mp},
which however did not separate between hard, jet and soft functions, such that their hard function is equal to the product of $H_{q\bar q} J_q J_\bq S_q$ when evaluated at fixed order.%
\footnote{An additional minor difference arises by normalizing by the total cross section $\sigma_T$ instead of the Born cross section $\sigma_0$.}
However, for the purpose of resummation, it is important to distinguish the hard function,
which describes physics at the high scale $\mu \sim Q$, from the jet and soft functions
which describe physics at the low scale $\mu \sim 1/b_T$. (In the context of TMD factorization,
the TMDFFs and soft function are often combined.) This will be addressed in \sec{literature}.
This separation was first achieved in \refcite{Moult:2018jzp}, and we closely follow their conventions.

\subsubsection{Higgs EEC}
\label{sec:fact_EEC_g}

We now discuss the extension of the EEC factorization to gluon-induced processes,
e.g.~the Higgs EEC in $e^+ e^- \to H \to g g + X$, which has not yet been given explicitly in the literature.
As in the quark case, one relates the EEC in the back-to-back limit to TMD factorization for identified hadron production,
$e^+ e^- \to h_1 h_2 + X$, which for gluon-induced processes reads
\begin{align} \label{eq:qT_fact_g}
 \frac{\df\sigma}{\df z_1 \df z_2 \df^2\qt} &
 =  2\born H_{\rho\sigma\rho'\sigma'}(Q, \mu) \int\!\df^2\bt \, e^{\img \qt \cdot \bt}
   \tilde D^{\rho\sigma}_{h_1 / g}\Bigl(z_1, \bt, \mu, \frac{\nu}{Q}\Bigr)
   \tilde D^{\rho'\sigma'}_{h_2 / g}\Bigl(z_2, \bt, \mu, \frac{\nu}{Q}\Bigr)
\nn\\&\qquad\times
   \tilde S_g(b_T, \mu, \nu)
 + \bigl[1 + \cO\bigl(q_T^2/Q^2\bigr)\bigr]
\,.\end{align}
To the best of our knowledge, this formula has not yet been explicitly given in the literature, but it follows immediately from the similar structure of TMD factorization at hadron colliders~\cite{Chiu:2012ir,Becher:2012yn,Echevarria:2015uaa}.
Similar to \eq{qT_fact_q}, $H$, $\tilde D_{h/g}$ and $\tilde S_g$ denote the hard, fragmentation and soft function, respectively.
For more details on the definition of the gluon fragmentation function, we refer to \refcite{Ebert:2020qef}.

The key difference between \eqs{qT_fact_q}{qT_fact_g} is the Lorentz structure of the hard function
and the fragmentation functions, reflecting the transverse polarization of the fragmenting gluons
in the factorization limit. Since the gluon fragmentation functions $\tilde D_{h / g}^{\rho\sigma}$ only depend on one Lorentz vector,
namely $b_\perp^\mu = (0, \bt, 0)$, their most general decomposition is given by
\begin{align} \label{eq:F_g_decomp}
 \tilde D_{h / g}^{\rho\sigma}(z,\bt)
 = \frac{g_\perp^{\rho\sigma}}{2} \tilde D_{h / g}(z,b_T)
 + \biggl(\frac{g_\perp^{\rho\sigma}}{2} - \frac{b_\perp^\rho b_\perp^\sigma}{b_\perp^2}\biggr) \tilde D'_{h/g}(z,b_T)
\,,\end{align}
where for brevity we suppressed the scales.

In this work, we will only be interested in the Higgs-initiated gluon EEC.
In this case, the scalar nature of the Higgs boson implies a trivial Lorentz structure of the hard function,
\begin{align}
 H^{\mu\nu\rho\sigma}(Q,\mu) = g_\perp^{\mu\rho} g_\perp^{\nu\sigma} H(Q, \mu)
\,.\end{align}
Inserting this into \eq{qT_fact_g} and using \eq{F_g_decomp}, we obtain
\begin{align} \label{eq:qT_fact_H}
 \frac{\df\sigma}{\df z_1 \df z_2 \df^2 \qt} &
 =\born  H(Q, \mu) \int\!\df^2\bt \, e^{\img \qt \cdot \bt}
   \Bigl[ \tilde D_{h_1 / g}\Bigl(z_1, \bt, \mu, \frac{\nu}{Q}\Bigr) \tilde D_{h_2 / g}\Bigl(z_2, \bt, \mu, \frac{\nu}{Q}\Bigr)
   \nn\\&\qquad
   + \tilde D'_{h_1 / g}\Bigl(z_1, \bt, \mu, \frac{\nu}{Q}\Bigr) \tilde D'_{h_2 / g}\Bigl(z_2, \bt, \mu, \frac{\nu}{Q}\Bigr) \Bigr]
   \tilde S_g(b_T, \mu, \nu)
\,,\end{align}
i.e.~we simply encounter the sum of two factorized expressions, one for the polarization-independent contribution
and one for the polarization-dependent contribution. (Note the factor $1/2$ from contracting \eq{F_g_decomp} with itself.)

We can now apply the same steps as in \sec{fact_EEC_q} to obtain the factorization formula for the Higgs EEC as
\begin{align} \label{eq:EEC_fact_thm_g}
 \frac{\df\sigma}{\df z} &
 = \frac{\born}{8} H(Q,\mu) \int_0^\infty \df (b_T Q)^2 \, J_0\bigl(b_T Q \sqrt{1-z}\bigr)
    \biggl[ J_g\Bigl(b_T, \mu, \frac{\nu}{Q}\Bigr)^2
         + J'_g\Bigl(b_T, \mu, \frac{\nu}{Q}\Bigr)^2 \biggr] \tilde S_g(b_T, \mu, \nu)
    \nn\\&\quad
    \times [ 1 + \cO(1-z) ]
\,.\end{align}
As before, the gluon jet functions are related to the matching coefficients of the gluon fragmentation functions,
\begin{align} \label{eq:TMDFFtoJ_g}
 J_g\Bigl(b_T, \mu, \frac{\nu}{Q}\Bigr) &= \sum_i \int_0^1\df z \, z \, \tilde\cK_{gi}\Bigl(z, b_T, \mu, \frac{\nu}{Q} \Bigr)
\,,\nn\\
 J'_g\Bigl(b_T, \mu, \frac{\nu}{Q}\Bigr) &= \sum_i \int_0^1\df z \, z \, \tilde\cK'_{gi}\Bigl(z, b_T, \mu, \frac{\nu}{Q} \Bigr)
\,,\end{align}
where $\tilde \cK_{gi}$ and $\tilde \cK'_{gi}$ are the matching kernels of $\tilde D_{h/g}$ and $\tilde D'_{h/g}$
onto the gluon fragmentation function $d_{h/g}$, with the matching taking the same form as \eq{F_matching}.
Since the polarized TMDFF $\cK'_{gi}$ starts at $\cO(\as)$, the same holds for the polarized jet function $J'_g$,
implying that it contributes to \eq{EEC_fact_thm_g} starting at $\cO(\as^2)$.
As in the quark case, the gluon jet functions $J_g$ and $J'_g$ depend on the chosen rapidity regulator,
and manifestly regulator-independent jet functions can be constructed as in \eq{calJdef},
see \sec{fact_EEC_q} for more details.

Note that our factorization theorem for the Higgs EEC disagrees with the statement in \refcite{Luo:2019hmp},
where the gluon jet function is claimed to be the linear combination $J_g + J'_g$.
In this case, the $J'_g$ would already contribute to the cross section at $\cO(\as)$.
However, by comparing our results for the Higgs EEC in the back-to-back limit with the fixed-order calculation of \refcite{Luo:2019nig},
we confirm that this is not the case, and find perfect agreement with the prediction from \eq{EEC_fact_thm_g}.

\subsection[Jet Function for the back-to-back limit at \texorpdfstring{N$^3$LO}{N3LO}]
           {\boldmath Jet function for the back-to-back limit at N$^3$LO}
\label{sec:jet_func}

Using \eqs{TMDFFtoJ}{TMDFFtoJ_g}, the EEC jet function can be easily obtained from the N$^3$LO result for the TMDFF calculated in our companion paper \cite{Ebert:2020qef}.
This result has been obtained using a recently developed method for the expansion of cross sections around the collinear limit \cite{Ebert:2020lxs}.

Throughout this paper we will present results computed with the exponential regulator \cite{Li:2016axz}. As a renormalization scheme we employ the one of \refcite{Chiu:2012ir} by relating the regulator $\tau$ to the inverse of the rapidity renormalization scale $\nu$. This is standard procedure for calculations in the exponential regulator and further details on it can be found in \refcite{Li:2016axz}.
The choice of this regulator is due to the fact that it allows to regulate rapidity divergences while only retaining information on the total momentum of the real radiation, which is at the basis of the framework of \refcite{Ebert:2020lxs} that we have employed for this calculation.
The rapidity-regulator independent combination $\cJ_i$ defined in \eq{calJdef}
can be obtained by combining our results with the N$^3$LO soft function computed in \refscite{Li:2016ctv,Ebert:2020lxs}.
While we only discuss results for $J_{q,g}$ in the following, for completeness we also provide $\cJ_{q,g}$ for quarks and gluons in the ancillary files.
Using $\cJ_i$, it is trivial to obtain the N$^3$LO jet function in \emph{any} rapidity regularization scheme
for which the soft function is known at the same order, which so far is only the case for the exponential regulator used here.

To present our results, we expand the jet functions as
\begin{align} \label{eq:J_expansion}
 J_i\Bigl(b_T, \mu, \frac{\nu}{Q}\Bigr) = \sum_{n=0}^\infty \Bigl[\frac{\as(\mu)}{4\pi}\Bigr]^n J_i^{(n)}(L_b, L_Q)
\,,\end{align}
where the $n$-th order coefficient $J_i^{(n)}$ only depends on the logarithms
\begin{align}
 L_b = \ln\frac{b_T^2 \mu^2}{b_0^2} \,, \quad L_Q = \ln\frac{\nu}{Q}
\,,\end{align}
with $b_0 = 2 e^{-\gamma_E}$.
The logarithmic structure of the  $J_i^{(n)}$ is fully encoded by the jet function RGEs,
which due to its definition in \eq{def_Jq} are identical to the RGEs of the TMDFF,
\begin{align} \label{eq:jet_RGEs}
 \frac{\df}{\df \ln \mu} \ln J_i(b_T,\mu,\nu/Q) &
 = \tilde\gamma_J^i(\mu,\nu/Q)
\,,\nn\\
 \frac{\df}{\df \ln \nu} \ln J_i(b_T,\mu,\nu/Q) &
 = -\frac12 \tilde\gamma_\nu^i(b_T,\mu)
\,,\end{align}
where the anomalous dimensions have the all-order expressions
\begin{align} \label{eq:jet_anom_dims}
 \tilde\gamma_J^i(\mu,\nu/Q) &
 = 2 \GammaC^i[\as(\mu)] \ln\frac{\nu}{Q} + \tilde\gamma_J^i[\as(\mu)]
\,,\nn\\
 \tilde\gamma^q_\nu(b_T,\mu) &=
 -4 \int_{b_0/b_T}^\mu \frac{\df\mu'}{\mu'} \, \GammaC^i[\as(\mu')] + \tilde\gamma_\nu^i[\as(b_0/b_T)]
\,.\end{align}
Here, $\GammaC^i(\as)$ is the cusp anomalous dimension,
$\tilde\gamma_J^i(\as)$ is the jet function noncusp anomalous dimension,
which is identical to the noncusp anomalous dimension $\gamma_B^i(\as)$ of the TMD beam function,
and $\tilde\gamma_\nu(\as)$ is the rapidity anomalous dimension.
Explicit results for these in our notation are collected in \refcite{Billis:2019vxg}.
For more details on these RGEs, see \sec{EEC_RGE}.

We have explicitly checked that our result for the jet function obeys \eqs{jet_RGEs}{jet_anom_dims}.
Solving these equations order-by-order in $\as$, one easily obtains the logarithmic structure of the $J_i^{(n)}(L_b, L_Q)$
in terms of the anomalous dimension and the constant piece of the jet function, which we define as
\begin{align}
 j_i^{(n)} \equiv J_i^{(n)}(L_b = L_Q = 0)
\,.\end{align}
In \app{EEC_jet_fixed_order}, we provide the full fixed-order structure of the $J_i^{(n)}$ through N$^3$LO,
and for brevity in the following we only present the constant terms $j_i^{(n)}$ through N$^3$LO.
The complete N$^3$LO jet functions are also provided as ancillary files with this submission.

\subsubsection{Quark jet function}
\label{sec:quark_jet_func}

The quark jet function has already been calculated using the exponential regulator in \refcite{Luo:2019hmp} at NLO and NNLO,
with which we find full agreement with their results. For completeness, we repeat these results,
\begin{align}\label{eq:J1_J2}
 j_q^{(0)} &= 1
\,,\nn\\
 j_q^{(1)} &= C_F (4 - 8 \zeta_2)
\,,\nn\\
 j_q^{(2)} &
 = C_F^2 \Bigl(\frac{139}{24} - 28 \zeta_2 - 74 \zeta_3 + 140 \zeta_4 \Bigr)
 + C_F C_A \Bigl( \frac{1549}{72} - \frac{178}{3} \zeta_2 + \frac{74}{3} \zeta_3  - 5 \zeta_4\Bigr)
 \nn\\&\quad
 + C_F n_f \Bigl(- \frac{149}{36} + \frac{28}{3} \zeta_2 + \frac43 \zeta_3 \Bigr)
\,.\end{align}
The new result of this paper is the three-loop coefficient, which is given by
\begin{align}\label{eq:j3q}
 j_q^{(3)} &
 = C_F^3 \left(-\frac{496}{3} \zeta_3^2 -\frac{17062}{9} \zeta_6 + 600
   \zeta_2 \zeta_3 +\frac{2008}{3}\zeta_5 +243 \zeta_4 +\frac{32}{3} \zeta_3 +\frac{22}{3}\zeta_2 +\frac{163}{4}\right)
   \nn\\&
 + C_F^2 C_A \left(\frac{184}{3}\zeta_3^2+\frac{2255}{9}\zeta_6- \frac{856}{9}\zeta_2\zeta_3+\frac{9620}{9}\zeta_5+\frac{46220}{27}\zeta_4-\frac{44224 }{27}\zeta_3-\frac{15004}{27}\zeta_2+\frac{24673}{216}\right)
 \nn\\& 	
   + C_F C_A^2 \left(\frac{56}{3}\zeta_3^2 +\frac{248}{9}\zeta_6 + 134 \zeta_2\zeta_3-\frac{2876}{3}\zeta_5 +\frac{19895}{108}\zeta_4+\frac{98269}{162}\zeta_3-\frac{107441}{162}\zeta_2+\frac{173785}{1944}\right)
    \nn\\&
    + C_F^2 n_f T_F \left( \frac{7976}{27}\zeta_2 +\frac{12584 \zeta_3}{27}-\frac{14188 \zeta_4}{27} -\frac{832}{9} \zeta_2\zeta_3 -\frac{2368 \zeta_5}{9}-\frac{3761}{27}\right)
 \nn\\&
   + C_F C_A n_f T_F \left(224 \zeta_5+ \frac{80}{3}\zeta_2\zeta_3-\frac{974 }{27}\zeta_4+\frac{928 \zeta_3}{81}+\frac{29900}{81}\zeta_2-\frac{16895}{243}\right)
    \nn\\&
   + C_F n_f^2 T_F^2
   \left(-\frac{352}{9}\zeta_2-\frac{5344 }{81}\zeta_3-\frac{448 \zeta_4}{27}+\frac{1586}{243}\right)
\,.\end{align}

\subsubsection{Gluon jet function}
\label{sec:gluon_jet_func}

In the gluon case, we have to consider both the polarization-independent jet function $J_g$
and the polarization-dependent jet function $J_g^\prime$.

At NLO and NNLO, the finite terms for $J_g$ can be obtained from the calculation of the unpolarized gluon TMDFFs of \refcite{Luo:2019bmw},
\begin{align}
 j^{(0)}_g &= 1
\,,\nn\\
 j^{(1)}_g &= C_A \left(\frac{65}{18}-8 \zeta_2\right) - \frac{5}{18} n_f
\,,\nn\\
 j^{(2)}_g &
 = C_A^2 \left(135 \zeta_4-\frac{176}{3}\zeta_3-\frac{727}{9}\zeta_2+\frac{2269}{81}\right)
   + C_A n_f \left(\frac{8}{3}\zeta_3+\frac{85}{9}\zeta_2-\frac{145}{162}\right)
   \nn\\&\quad
   + C_F n_f \left(8 \zeta_3-\frac{49}{6}\right)
   - \frac{14}{81} n_f^2
\,.\end{align}
The new result at three loops can be obtained from our calculation of the TMDFF in \refcite{Ebert:2020qef}. We obtain
\begin{align}\label{eq:j3g}
 j^{(3)}_g &
 = C_A^3 \Bigl(\frac{887378}{3645} -\frac{7091}{6}\zeta_2 -\frac{52064}{45}\zeta_3 + \frac{101995}{54}\zeta_4 + \frac{8540}{9}\zeta_5 +\frac{2228}{3} \zeta_2 \zeta_3 - \frac{4853}{3}\zeta_6 -\frac{256}{3} \zeta_3^2 \Bigr)
 \nn\\&
 + C_A^2 n_f \Bigl(\frac{111563}{2430} + \frac{6937}{27}\zeta_2 + \frac{77459}{270}\zeta_3 - \frac{16631}{108}\zeta_4 - \frac{2060}{9}\zeta_5 -\frac{58}{3} \zeta_2 \zeta_3  \Bigr)
 \nn\\&
 +C_A C_F n_f \Bigl(\frac{320843}{1620} + \frac{1168}{9}\zeta_2 + \frac{3254}{45}\zeta_3 - \frac{8}{9}\zeta_4 + 120 \zeta_5 -\frac{352}{3} \zeta_2 \zeta_3 \Bigr)
 \nn\\&
 +C_A n_f^2 \Bigl(-\frac{59131}{4860} -\frac{349}{27}\zeta_2 - \frac{94}{45}\zeta_3 - \frac{100}{9}\zeta_4\Bigr)
 \nn\\&
 +C_F^2 n_f \Bigl(\frac{331}{18} +  \frac{148}{3}\zeta_3 - 80 \zeta_5\Bigr)
 +C_F n_f^2 \Bigl(\frac{1624}{81} - \frac{44}{3}\zeta_3 \Bigr)
 +n_f^3 \Bigl(\frac{494}{729} - \frac{32}{27}\zeta_3 \Bigr)
\,.\end{align}
The polarization-dependent jet function $J'_g$ starts at $\cO(\as)$, but for Higgs production only interferes with itself, see \eq{EEC_fact_thm_g}.
Hence, it is sufficient to know $J'_g$ at $\cO(\as^2)$ to obtain the Higgs EEC at N$^3$LO.
For completeness, we report their results from \refcite{Luo:2019bmw},
\begin{align}\label{eq:Jprime}
 j^{\prime(0)} &= 0
\,,\nn\\
 j^{\prime(1)}_g &= \frac{C_A}{3}-\frac{n_f}{3}
\,,\nn\\
 j^{\prime(2)}_g &= C_A^2 \left(-\frac{8\zeta_2}{3}+\frac{107}{27}\right)
                  + C_A n_f \left(\frac{8 \zeta_2}{3}-\frac{163}{27}\right)
                  + 2 C_F n_f + \frac{2}{27} n_f^2
\,.\end{align}

\subsection[The EEC in the back-to-back limit at \texorpdfstring{N$^3$LO}{N3LO}]
           {\boldmath The EEC in the back-to-back limit at N$^3$LO}
\label{sec:EEC_back_to_back}

Combining our result with the hard function from \refcite{Gehrmann:2010ue} and the soft function from \refcite{Li:2016ctv},
we have all ingredients to obtain the EEC in the $z\to1$ limit at N$^3$LO.
The Bessel integral in \eqs{EEC_fact_thm_q}{EEC_fact_thm_g} can be easily evaluated analytically, see \app{EEC_Bessel}.

To present our results, we expand the EEC in the back-to-back limit as
\begin{align}\label{eq:EECN3LOFO}
\dso
 =\born  \sum_{n=0}^\infty \Bigl[\frac{\as(\mu)}{4\pi}\Bigr]^n \frac{\df\bar\sigma^{(n)}(L_h)}{\df z}
\,,\end{align}
where we have divided out the Born partonic cross section $\born$.
In \eq{EECN3LOFO}, $L_h = \ln(Q/\mu)$ is the only remaining logarithm of $\mu$, and its structure is entirely governed by the $\beta$ function.
For brevity, we only report the nonlogarithmic terms (which is the same as the entire result for the choice $\mu = Q$), but provide the full result in an ancillary file with this submission.

\subsubsection{Quark EEC}
\label{sec:EEC_q}

The results through NNLO were already given in \refcite{Luo:2019hmp}, with which we fully agree, and which are repeated here for completeness:
\begin{align} \label{eq:EEC_z1_NLO_NNLO}
 \frac{\df \bar\sigma^{(0)}}{\df z} &
 = \frac{1}{2} \delta(\zb)
\,,\nn\\
 \frac{1}{C_F} \frac{\df \bar\sigma^{(1)}}{\df z} &
 = - 2 \cL_1(\zb) - 3 \cL_0(\zb) - (4 + 2 \zeta_2) \delta(\zb)
\,,\nn\\
 \frac{1}{C_F} \frac{\df \bar\sigma^{(2)}}{\df z} &
 = 4 C_F \cL_3(\zb)
 + \cL_2(\zb) \Bigl( 18 C_F + \frac{22}{3} C_A - \frac{4}{3} n_f \Bigr)
 \nn\\&\quad
 + \cL_1(\zb) \Bigl[ C_F (34 + 8 \zeta_2) + C_A \Bigl(-\frac{35}{9} + 4 \zeta_2\Bigr) + \frac29 n_f \Bigr]
 \nn\\&\quad
 + \cL_0(\zb) \Bigl[ C_F \Bigl( \frac{45}{2} + 24 \zeta_2 - 8 \zeta_3\Bigr) + C_A \Bigl(-\frac{35}{2} + 22 \zeta_2 + 12 \zeta_3 \Bigr) + n_f (3 - 4 \zeta_2) \Bigr]
 \nn\\&\quad
 + \delta(\zb) \Bigl[
     C_F \Bigl( \frac{41}{3} + 49 \zeta_2 - 80 \zeta_3 + 48 \zeta_4 \Bigr)
   + C_A \Bigl( -\frac{382}{9} - \frac{104}{9} \zeta_2 + \frac{182}{3} \zeta_3 - 8 \zeta_4 \Bigr)
   \nn\\&\hspace{1.5cm}
   + n_f \Bigl( \frac{58}{9} + \frac{8}{9} \zeta_2 + \frac{4}{3} \zeta_3 \Bigr)
  \Bigr]
\,.\end{align}
Here, we introduced the standard plus distributions
\begin{align}\label{eq:Lzbdef}
 \cL_n(\zb) = \Bigl[ \frac{\ln^n \zb}{\zb} \Bigr]_+ \,, \quad \zb \equiv 1-z
\,.\end{align}
The new result at N$^3$LO reads
{\allowdisplaybreaks
\begin{align} \label{eq:EEC_z1_N3LO_ee}
 \frac{1}{C_F} \frac{\df \bar\sigma^{(3)}}{\df z} &
 = -4 C_F^2 \cL_5(\zb)
 \nn\\&\quad
  + \cL_4(\zb) \Bigl[-30 C_F^2 - \frac{220}{9} C_F C_A  + \frac{40}{9} C_F n_f \Bigr]
 \nn\\&\quad
  + \cL_3(\zb) \Bigl[ C_F^2 (-16 \zeta_2-104) +\frac{88}{9} C_F n_f + C_F C_A \Bigl(-16 \zeta_2-\frac{388}{9}\Bigr)
    \nn\\&\hspace{2cm}
    - \frac{242}{9} C_A^2 + \frac{88}{9} C_A n_f - \frac{8}{9} n_f^2 \Bigr]
 \nn\\&\quad
  + \cL_2(\zb) \Bigl[
     C_F^2 (-144 \zeta_2-16 \zeta_3-189)
   + C_F C_A \Bigl(-\frac{592}{3}\zeta_2-72 \zeta_3+\frac{244}{3}\Bigr)
   \nn\\&\hspace{2cm}
   + C_F n_f \Bigl(\frac{88}{3}\zeta_2-\frac{40}{3}\Bigr)
   + C_A^2\Bigl(\frac{2471}{27}-\frac{88}{3}\zeta_2 \Bigr)
   \nn\\&\hspace{2cm}
   + C_A n_f \Bigl(\frac{16}{3}\zeta_2 - \frac{760}{27}\Bigr)
   + \frac{44n_f^2}{27}
  \Bigr]
 \nn\\& \quad
  + \cL_1(\zb) \Bigl[
    C_F^2 \Bigl(-\frac{542}{3} -412 \zeta_2 + 224 \zeta_3-192 \zeta_4\Bigr)
    \nn\\&\hspace{2cm}
    + C_F C_A \Bigl(-\frac{2900}{9}\zeta_2-\frac{1688}{3}\zeta_3-8 \zeta_4+\frac{3797}{9}\Bigr)
    \nn\\&\hspace{2cm}
    + C_F n_f \Bigl(\frac{536}{9}\zeta_2+\frac{32}{3}\zeta_3-\frac{479}{9}\Bigr)
    + C_A^2 \Bigl(-\frac{916}{9}\zeta_2-44 \zeta_4-\frac{2354}{81}\Bigr)
    \nn\\&\hspace{2cm}
    + C_A n_f\Bigl(\frac{448}{9}\zeta_2 +16 \zeta_3-\frac{380}{81}\Bigr)
    + n_f^2\Bigl(\frac{124}{81}-\frac{16}{3}\zeta_2\Bigr)
   \Bigr]
 \nn\\&\quad
  + \cL_0(\zb) \Bigl[
      C_F^2 \Bigl(64 \zeta_3 \zeta_2-402\zeta_2+332 \zeta_3-552 \zeta_4+48 \zeta_5-\frac{169}{2}\Bigr)
    \nn\\&\hspace{2cm}
    + C_F C_A \Bigl(-128\zeta_3 \zeta_2+\frac{212}{3}\zeta_2 -\frac{2812}{9}\zeta_3 -\frac{1342}{3}\zeta_4 - 120 \zeta_5+\frac{3358}{9}\Bigr)
    \nn\\&\hspace{2cm}
    + C_F n_f \Bigl(-\frac{20}{3}\zeta_2 -\frac{296}{9}\zeta_3 + \frac{244}{3}\zeta_4 - \frac{623}{18}\Bigr)
    \nn\\&\hspace{2cm}
    + C_A^2 \Bigl(\frac{4420}{9}\zeta_2 -\frac{560}{9}\zeta_3-\frac{326}{3}\zeta_4 - 40 \zeta_5-\frac{4241}{27}\Bigr)
    \nn\\&\hspace{2cm}
    + C_A n_f\Bigl(-\frac{1508}{9} \zeta_2 +\frac{184}{9}\zeta_3 + \frac{56}{3}\zeta_4 + \frac{1414}{27}\Bigr)
    \nn\\&\hspace{2cm}
    + n_f^2 \Bigl(\frac{112}{9}\zeta_2 + \frac{16}{9}\zeta_3 - \frac{98}{27}\Bigr)
    \Bigr]
 \nn\\&\quad
 + \delta(\zb) \Bigl[ C_F^2 \Bigl(-\frac{337}{3} -\frac{1049}{3}\zeta_2 +\frac{530}{3} \zeta_3 + 512 \zeta_2 \zeta_3 - 64 \zeta_3^2 -1396 \zeta_4+\frac{3136 }{3} \zeta_5-672 \zeta_6\Bigr)
    \nn\\&\hspace{1.5cm}
    + C_F C_A \Bigl(\frac{10169}{27} +\frac{2729}{3}\zeta_2 -\frac{22070}{9}\zeta_3 +\frac{2176 \zeta_4}{9}+ 528 \zeta_5 + 22 \zeta_6 -288 \zeta_2 \zeta_3 + 64 \zeta_3^2\Bigr)
    \nn\\&\hspace{1.5cm}
    + C_F n_f \Bigl(-\frac{148}{27} -\frac{985 \zeta_2}{9}+\frac{3340 \zeta_3}{9}+\frac{58 \zeta_4}{9}-\frac{368 \zeta_5}{3} -\frac{224}{3} \zeta_2 \zeta_3 \Bigr)
    \nn\\&\hspace{1.5cm}
    + C_A^2 \Bigl(-\frac{55504}{81} -\frac{3968}{81}\zeta_2 +\frac{39337}{27}\zeta_3 +\frac{3815}{18}\zeta_4 - \frac{2720}{3}\zeta_5 -\frac{700}{3}\zeta_2 \zeta_3 + 59 \zeta_6  -56 \zeta_3^2 \Bigr)
    \nn\\&\hspace{1.5cm}
    + C_A n_f \Bigl(\frac{15626}{81} - \frac{3326}{81}\zeta_2 - \frac{3788}{27}\zeta_3 - \frac{290}{9}\zeta_4 + 80\zeta_5 + 72 \zeta_2 \zeta_3 \Bigr)
    \nn\\&\hspace{1.5cm}
    + n_f^2 \Bigl(-\frac{1048}{81} + \frac{616}{81}\zeta_2 - \frac{464}{27}\zeta_3 - \frac{16}{9}\zeta_4 \Bigr)
    \nn\\&\hspace{1.5cm}
    + N_{F,V} \frac{d_{abc}d^{abc}}{N_r} (2+5 \zeta_2+\frac{7}{3} \zeta_3-\frac{\zeta_4}{2}-\frac{40}{3} \zeta_5)
\Bigr]
\,.\end{align}
}%
Here, following the notation of \refcite{Gehrmann:2010ue} the factor $N_{F,V}$ arises from virtual diagrams where the exchanged vector boson couples to a closed quark loop, and hence this contribution is not proportional to the charges of the Born process.
For the simplest case of photon exchange, it is given by $N_{F,V} = (\sum_f e_f)/e_q$, where $e_q$ is the flavor of the external quark. Note that $d_{abc}d^{abc}= (N_c^2-4)(N_c^2-1)/N_c$ and $N_r$ is the dimension of the fundamental representation, hence $N_r=3$ in QCD.

\subsubsection{Gluon EEC}
\label{sec:EEC_g}

The singular structure of the Higgs EEC in the back-to-back limit is given by
\begin{align}\label{eq:EEC_z1_NNLO_H}
\frac{\df \bar\sigma_H^{(0)}}{\df z}&=\frac{1}{2} \delta (\zb)
\nn\\
\frac{\df \bar\sigma_H^{(1)}}{\df z}&=-2 C_A
   \cL_1(\zb) -\cL_0(\zb)\left[\frac{11}{3}C_A-\frac{2}{3}n_f\right]+ \delta (\zb)\left[C_A \left(\frac{65}{18}-2  \zeta_2\right)-n_f\frac{5}{18}\right]
\nn\\
\frac{\df \bar\sigma_H^{(2)}}{\df z}&=4 C_A^2\cL_3(\zb)+\cL_2(\zb)\left[\frac{88}{3}C_A^2-\frac{16}{3}C_A n_f\right]+\cL_1(\zb) \left[C_A^2\left(12 \zeta_2
   +11\right)-\frac{34}{3}n_f C_A+\frac{4}{3}n_f^2\right]
\nn\\&+
   \cL_0 \left[C_A^2\left(4 \zeta_3 +\frac{154}{3} \zeta_2-\frac{907}{18}\right)
   +C_A n_f \left(\frac{233}{18}-\frac{28}{3}  \zeta_2\right)+2 C_F n_f -\frac{5}{9}n_f^2\right]
\nn\\&+
	\delta (\zb)\left[C_A^2\left(40 \zeta_4 -44 \zeta_3 -\frac{65}{3} \zeta_2 +\frac{17515}{216}\right)+C_An_f \left(-8 \zeta_3
   +\frac{5}{3} \zeta_2 -\frac{1657}{108}\right)
\right.\nn\\&\left.\hspace{1.5cm}  
	-\frac{17 }{216}n_f^2+C_F n_f \left(16  \zeta_3-\frac{58}{3}\right)\right]\,.
\end{align}
The logarithmic structure matches the fixed order calculation of \refcite{Luo:2019nig}, while to the best of our knowledge, the $\delta(\zb)$ term has not appeared in the literature before.
Note that the polarized jet function $J'_g$ appears for the first time in the $\delta(\zb)$ coefficient at $\cO(\alpha_s^2)$,
as it vanishes at tree level and interferes only with itself.

For the singular structure at N$^3$LO we find
{\allowdisplaybreaks
\begin{align}\label{eq:EEC_z1_N3LO_H}
\frac{\df \bar\sigma_H^{(3)}}{\df z}&=-4C_A^3\cL_5(\zb)
\nn\\&-\cL_4(\zb)\left[\frac{550}{9}C_A^3 -\frac{100}{9}C_A^2 n_f\right]
\nn\\&+
\cL_3(\zb) \left[C_A^3 \left(-32 \zeta_2-\frac{4630}{27}\right)+\frac{2252}{27}C_A^2 n_f-\frac{232}{27}C_An_f^2\right]
\nn\\&+
   \cL_2(\zb) \left[ C_A^3\left(-88 \zeta_3-\frac{1232}{3} \zeta_2 +\frac{9589}{27}\right)+C_A^2 n_f \left(\frac{224}{3}  \zeta_2-\frac{731}{27}\right)
   -\frac{482 }{27}C_A n_f^2
\right.\nn\\&\left.\hspace{1.5cm}
   -16 C_A C_F n_f+\frac{16}{9}n_f^3\right]
\nn\\&+
\cL_1(\zb) \left[C_A^3\left(-244 \zeta_4-\frac{880}{3} \zeta_3 -\frac{2200}{3} \zeta_2 +\frac{12719}{54}\right)+ C_A^2n_f \left(\frac{400}{3} \zeta_3
   +\frac{976}{3} \zeta_2 -\frac{2173}{9}\right)
\right.\nn\\&\left.\hspace{1.5cm}+
	C_A n_f^2\left(\frac{811}{18}-32  \zeta_2\right)+C_A C_F n_f
   \left(\frac{133}{3}-80  \zeta_3\right)+C_F n_f^2\frac{28}{3}-\frac{40}{27}n_f^3\right]
\nn\\&+
   \cL_0(\zb) \left[C_A^3\left(-64 \zeta_2 \zeta_3 -112 \zeta_5 -1232 \zeta_4 +\frac{1280}{9} \zeta_3 +\frac{30232}{27} \zeta_2 -\frac{81466}{81}\right)
\right.\nn\\&\left.\hspace{1.5cm}
   +C_A^2 n_f \left(224 \zeta_4 +\frac{436}{3} \zeta_3 -\frac{8720}{27} \zeta_2 +\frac{18079}{54}\right) +C_A n_f^2\left(-\frac{272}{9} \zeta_3 +\frac{520}{27} \zeta_2 -26\right)
\right.\nn\\&\left.\hspace{1.5cm}
   +C_A C_F n_f \left(-176 \zeta_3 -28 \zeta_2
   +\frac{4261}{18}\right)-C_F^2 n_f+C_F n_f^2 \left(32  \zeta_3-\frac{356}{9}\right) -\frac{17}{81}n_f^3 \right]+
\nn\\&+
   \delta(\zb)\left[ C_A^3\left(-56 \zeta_3^2 -591 \zeta_6 +68 \zeta_2 \zeta_3 +\frac{2128}{3} \zeta_5
   -\frac{916}{3} \zeta_4 -\frac{271522}{135} \zeta_3 -\frac{50077}{54} \zeta_2 +\frac{14650931}{7290}\right)
\right.\nn\\&\left.\hspace{1.5cm}+
    C_A^2n_f\left(\frac{166}{3} \zeta_2 \zeta_3 -\frac{316}{3} \zeta_5 +\frac{10345}{36} \zeta_4 +\frac{27833}{90} \zeta_3 +\frac{8254}{27}
   \zeta_2 -\frac{3381919}{4860}\right)
\right.\nn\\&\left.\hspace{1.5cm}+
   C_A C_F  n_f \left(-\frac{208}{3} \zeta_2 \zeta_3 +200 \zeta_5 +4 \zeta_4 +\frac{15254}{45}
   \zeta_3 +\frac{449}{3} \zeta_2 -\frac{189683}{270}\right)
\right.\nn\\&\left.\hspace{1.5cm}+
	C_F n_f^2 \left(-\frac{8}{9} \zeta_4 -\frac{176}{3} \zeta_3 -\frac{92}{9} \zeta_2+\frac{8983}{108}\right)
	+C_F^2 n_f \left(-240 \zeta_5 +148 \zeta_3 +\frac{313}{6}\right)
\right.\nn\\&\left.\hspace{1.5cm}+
	C_An_f^2\left(-\frac{284}{9} \zeta_4 -\frac{314}{45} \zeta_3 -\frac{620}{27} \zeta_2+\frac{105647}{2430}\right) 
	+n_f^3 \left(\frac{32}{27} \zeta_3 +\frac{511}{729}\right)\right]
\,.\end{align}
}

\subsection[The maximally transcendental limit from \texorpdfstring{$\cN=4$}{N = 4} sYM]
           {\boldmath The maximally transcendental limit from $\cN = 4$ sYM}
\label{sec:maximum_transcendentality}

It is conjectured that the leading transcendental term of the EEC in QCD (with $C_F,C_A \to N_c$) in the back-to-back limit
is given by the EEC in the back-to-back limit in maximally supersymmetric Yang-Mills theory ($\cN=4$ SYM)~\cite{Belitsky:2013ofa}.
This relation has been observed to hold up to $\cO(\alpha_s^2)$ \cite{Dixon:2018qgp,Luo:2019nig,Luo:2019hmp,Dixon:2019uzg} and it is known not to hold beyond leading power \cite{Moult:2019vou,Henn:2019gkr}.
Here, we validate this conjecture at $\cO(\as^3)$, which also provides us with an independent check of the leading transcendental limit of the three-loop jet functions.

In $\cN=4$ SYM, the back-to-back limit of the EEC is given by a (simplified) conformal version of the factorization formula in \eq{EEC_fact_thm_q} \cite{Belitsky:2013ofa,Korchemsky:2019nzm,Kologlu:2019mfz},
\begin{align} \label{eq:EEC_fact_SYM}
 \lim_{z\to1} \!\frac{\df \sigma^{\cN=4}}{\df z} &
 = \frac{\born}{8} \frac{H(a)}{\zb} \int_0^\infty\!\! \df b \, b J_0(b)
   \exp\left[-\frac12 \GammaC(a) \ln^2\left(\frac{b^2}{\zb b_0^2}\right)-\Gamma(a)\ln\left(\frac{b^2}{\zb b_0^2}\right) \right]
.\end{align}
Here, $\zb \equiv 1-z$, and $\GammaC(a)$ and $\Gamma(a)$ are the cusp and the collinear anomalous dimensions \cite{Korchemsky:1987wg,Bern:2005iz,Henn:2019swt,Dixon:2008gr,Dixon:2017nat}.
In \refcite{Korchemsky:2019nzm} it has been shown that the boundary function $H(a)$ can be obtained from the OPE coefficients of twist-two operators at large spin up to 3 loops~\cite{Eden:2012rr,Alday:2013cwa} and it reads 
\beq\label{eq:KorchemskyH}
	H(a) = 1 - \zeta_2  a + 5 \zeta_4 a^2 - \left(\frac{17}{12}\zeta_3^2 + \frac{591}{32}\zeta_6\right)a^3 + \cO(a^4)\,.
\eeq
While $H(a)$ is sometimes referred as the hard function of the back-to-back asymptotic in $\cN=4$ SYM,
it is important to notice that it is neither the $\cN=4$ analog nor the maximally transcendental part
of the QCD hard function $H(Q,\mu)$ which appears in the factorization theorem in \eq{EEC_fact_thm_q} (or in \eq{EEC_fact_thm_g} for that matter).
The reason lies in the fact that in QCD the running of the coupling forces a distinction between
the constants of the $\delta(1-z)$ term due to hard, collinear or soft corrections.
However, in $\cN=4$, because of conformal invariance, there is no reason for such separations
and therefore the non-logarithmic enhanced corrections get combined into a single term, namely $H(a)$.

Combining \eqs{EEC_fact_SYM}{KorchemskyH}, we obtain the contact term of the $z=1$ endpoint in $\cN=4$ up to three loops \cite{Korchemsky:2019nzm,Kologlu:2019mfz},
\begin{align} \label{eq:N1Prediction}
 \frac{1}{\born}\frac{\df \sigma^{\cN=4}}{\df z} \bigg|_{\delta(1-z)}
 = \frac12 \delta(1-z) \Bigl[& 1 -4\zeta_2 \left(N_c\frac{\alpha_s}{4\pi}\right)
   + 80 \zeta_4 \left(N_c\frac{\alpha_s}{4\pi}\right)^2
   \nn\\&
   - (112\zeta_3^2 + 1182\zeta_6) \left(N_c\frac{\alpha_s}{4\pi}\right)^3
   + \cO(\as^4) \Bigr].
\end{align}
If the principle of maximal transcendentality holds at this order, \eq{N1Prediction} should predict the leading transcendental terms
of the $\delta(1-z)$ coefficient of the EEC in QCD through $\cO(\alpha_s^3)$.%
\footnote{Note that starting at $\cO(\alpha_s^3)$ there is a leading transcendental contribution
to the $\delta(1-z)$ term from the Fourier transform of the cusp, see appendix~\ref{app:EEC_Bessel}.
Therefore, the boundary terms do not agree between $b_T$ space and $z$ even at leading transcendental weight.}

We find perfect agreement between \eq{N1Prediction} and the leading transcendental terms
of the quark and gluon EEC in \eqs{EEC_z1_N3LO_ee}{EEC_z1_N3LO_H}, confirming that the conjectured
principle of maximal transcendentality holds for the EEC in the back-to-back limit through $\cO(\alpha_s^3)$.

Conversely, if we instead assume that \eq{N1Prediction} predicts the maximal transcendental limit of the EEC in QCD,
we can use it to predict the leading transcendental term of the EEC jet function at $\cO(\as^3)$,
as the required hard and soft functions are already known at this order~\cite{Gehrmann:2010ue, Li:2016ctv}.
We obtain
\begin{align}
 j_{q,g}^{(3)~\rm{l.t.}} = -N_c^3 \Bigl( \frac{256}{3} \zeta_3^2 + \frac{4853}{3} \zeta_6\Bigr)
\,,\end{align}
which precisely agrees with the leading transcendental limit of the quark and gluon jet functions given in \eqs{j3q}{j3g}.
Since the jet functions are obtained from a weighted integral of the corresponding TMDFF,
summed over all contributing partonic channels, this also provides a remarkable cross check
on the N$^3$LO TMDFF matching kernels calculated in the companion paper~\cite{Ebert:2020qef}.

\section{\boldmath Sum rules and the \moment of the EEC at \texorpdfstring{$\cO(\alpha_s^3)$}{O(as3)}}
\label{sec:sum_rules}

An interesting property of the EEC is that it obeys the sum rules~\cite{Korchemsky:2019nzm,Dixon:2019uzg,Kologlu:2019mfz}
\begin{align}\label{eq:sumruleEEC}
	\int_0^1 \df z \frac{\df \sigma}{\df z} &= \sigma \\
	\int_0^1 \df z z \frac{\df \sigma}{\df z} &= \int_0^1 \df z (1-z) \frac{\df \sigma}{\df z} = \frac{1}{2}\sigma\,,
\end{align}
where $\sigma$ is the inclusive cross section for $e^+ e^- \to $ hadrons (or Higgs decay to hadrons), which is known up to $\cO(\alpha_s^4)$ \cite{Baikov:2012er,Baikov:2012zn,Herzog:2017dtz}. These sum rules are due to energy and momentum conservation, respectively.
The first sum rule can be derived by noting that 
\begin{align}
 \sum_{a,b} \int \df\sigma_{V\to a+b+X} = \sigma
\,,\qquad
 \sum_{a,b}\frac{E_aE_b}{Q^2} = \left(\sum_{a}\frac{E_a}{Q}\right)\left(\sum_{b}\frac{E_b}{Q}\right) = 1\,.
\end{align}
The second follows from
\beq
	p_a \cdot p_{b} = E_a E_b - E_a E_b \cos \chi_{a,b}\,,
\eeq
such that 
\begin{align}
 \sum_{a,b} \frac{E_aE_b}{Q^2} \frac{(1 - \cos \chi_{a,b})}{2}  =\left(\sum_{a}\frac{p_a^\mu}{Q}\right)\left(\sum_{b}\frac{p_{b\mu}}{Q}\right)\,,
\end{align}
where $\sum_i p_i^\mu = q^\mu = (Q,0,0,0)$ in the rest frame of the source.\footnote{Note that the choice of frame is already implicitly taken in defining the EEC observable in terms of the angle $\chi_{a,b}$ which is clearly frame dependent. An alternative, and equivalent, way of defining the observable in a frame independent way is to let $ z = \frac{p_a \cdot p_b}{2p_a\cdot q p_b \cdot q}$. It is easy to check that, in the rest frame of the source, this gives back the definition in \eq{z}.}
The presence of these sum rules puts interesting constraints on the EEC distribution and allows one to use information about one region to extract nontrivial information on the distribution away from that region~\cite{Korchemsky:2019nzm}.
In particular it is useful to divide the EEC distribution in different terms \cite{Korchemsky:2019nzm,Dixon:2019uzg}
\begin{align}
	\frac{\df \sigma}{\df z} &= V_0 \delta(z) + \left[\frac{\phi_0(z)}{z}\right]_+ 
	\quad+\quad
	V_1 \delta(1-z) + \left[\frac{\phi_1(z)}{1-z}\right]_+
	\quad+\quad
	\EECreg\,.
\end{align}
This isolates the endpoint singular terms with respect to the regular part of the distribution, such that $\left.\frac{\df \sigma}{\df z}\right|_{\text{reg.}}$ is finite as $z\to 0$ and $z\to 1$ and the plus prescription for $\phi_0$ is taken to act at $z=0$, while that for $\phi_1$ at $z=1$.
The functional form of the leading term of the EEC in the $z\to 0$ and $z \to 1$ limit in QCD is known at all orders due to the factorization theorems presented in \refcite{Dixon:2019uzg} and \refcite{Moult:2018jzp}.
In particular, order by order in $\alpha_s$ one finds
\begin{align} \label{eq:EEC_split}
	\phi_0(z) &=\born \sum_{n}\left(\frac{\alpha_s}{4\pi}\right)^n \sum_{m=0}^{n-1} c^{(n,m)}_{0} \log^m z\nn \\
	\phi_1(z) &=\born \sum_{n}\left(\frac{\alpha_s}{4\pi}\right)^n \sum_{m=0}^{2n-1} c^{(n,m)}_{1}\log^m (1-z)\,,
\end{align}
where all the coefficients $c_{n,m}^{(k)}$ with $n\leq 3$ are known \cite{Dixon:2019uzg,Moult:2018jzp}. In particular, the coefficient $c^{(3,m)}_{1}$ is the coefficient of $\cL_m(\zb)$ in \eqs{EEC_z1_N3LO_ee}{EEC_z1_N3LO_H} for $e^+e^-$ and Higgs, respectively.
Expanding the total cross section and the boundary constants as  
\begin{align}\label{eq:constDef}
	\sigma= \born\sum_n \left(\frac{\alpha_s}{4\pi}\right)^n R^{(n)}\,, \quad V_0= \born\sum_n \left(\frac{\alpha_s}{4\pi}\right)^n V_0^{(n)}\,, \quad V_1= \born\sum_n \left(\frac{\alpha_s}{4\pi}\right)^n V_1^{(n)}\,,
\end{align}
we can make use of the sum rules and obtain relations order by order in $\alpha_s$. We can write the relations at $\cO(\alpha_s^n)$ in compact form as 
\begin{align}
	R^{(n)} &= V_0^{(n)} + V_1^{(n)} +  \frac{1}{\born}\int \df z \EECregas{n} \label{eq:sumRuleNorderN} \\
	\frac{R^{(n)}}{2} &= V_1^{(n)} + \sum_{m=0}^{2n-1} (-1)^m m! \bigl[c^{(n,m)}_{0} - c^{(n,m)}_{1}\bigr] + \frac{1}{\born} \int \df z \, z \EECregas{n}	\label{eq:sumRulez1} \\
	\frac{R^{(n)}}{2} &= V_0^{(n)} + \sum_{m=0}^{2n-1} (-1)^m m! \bigl[c^{(n,m)}_{1} - c^{(n,m)}_{0}\bigr] +  \frac{1}{\born}\int \df z \, (1-z) \EECregas{n}\label{eq:sumRulez0}\,.
\end{align}
In \app{sumruleIngredients} we collect the expressions for the terms entering \eq{sumRuleNorderN}, \eq{sumRulez1} and \eq{sumRulez0}.
The different ingredients can be obtained with wildly different techniques and, as it is often the case, some are easier to obtain than others. Therefore, the power of these sum rules lies in the ability of extracting information about the endpoints from the bulk of the distribution and vice versa.
This idea has been applied in \refcite{Korchemsky:2019nzm} to obtain the endpoint behaviors of the EEC in $\cN=4$ up to three loops.%
\footnote{The contact terms of the EEC in $\cN=4$ were also obtained in \refcite{Kologlu:2019mfz}.}
In QCD, it has been applied in \refcite{Dixon:2019uzg} to extract $V_0^{(2)}$, i.e.\ the two loop $\delta(z)$ coefficient, using the two loop $\delta(1-z)$ coefficient $V_1^{(2)}$ from \refscite{Luo:2019hmp,Luo:2019bmw} and the regular part of the distribution $\EECregas{2}$ from the fixed order calculation of \refscite{Dixon:2018qgp,Luo:2019nig} both for the EEC in $e^+e^-$ as well as in gluon-induced Higgs decay.

While all the ingredients are known in QCD at $\cO(\alpha_s^2)$, much less is known at $\cO(\alpha_s^3)$.
In particular, only the logarithmic structure at the endpoint, i.e.\ the $c^{(3,m)}_{1}$ and $c^{(3,m)}_{0}$ coefficients, is known~\cite{Dixon:2019uzg,Moult:2018jzp},
but neither the regular part of the distribution nor the $\delta(z),\delta(1-z)$ terms are known.
In this work we have calculated the EEC jet function at $\cO(\alpha_s^3)$, which is the last missing ingredient to obtain $V_1^{(3)}$, i.e.\ the three-loop coefficient of $\delta(1-z)$, which we have presented in \eqs{EEC_z1_N3LO_ee}{EEC_z1_N3LO_H} for $e^+e^-$ and Higgs, respectively.
Combining our new results with \eq{sumRulez1}, we obtain the \moment of the EEC in the bulk at $\cO(\as^3)$.

For $e^+e^-$ we obtain 
\begin{align}\label{eq:sumrule_ee}
\frac{1}{\born} &\int_0^1\!\! \df z \, z \EECregaslab{3}{e^+e^-}
\nn\\&
 = C_F^3 \Bigl(64 \zeta_3^2+672 \zeta_6-496 \zeta_2 \zeta_3-\frac{3616}{3}\zeta_5+\frac{4778
   \zeta_4}{3}-210 \zeta_3+\frac{99397}{216}\zeta_2 - \frac{3809015}{7776}\Bigr)
 \nn\\&
 + C_F n_f^2 \Bigl(\frac{16}{9}\zeta_4+\frac{56}{27}\zeta_3+\frac{3094}{405}\zeta_2+\frac{156437}{13500}\Bigr)
\nn \\ &+
   C_F C_A^2 \Bigl(56 \zeta_3^2 -59 \zeta_6 +\frac{628}{3} \zeta_2 \zeta_3+\frac{2212}{3}\zeta_5-\frac{1768}{9}\zeta_4-\frac{438601}{270}\zeta_3
   + \frac{7930931}{8100}\zeta_2-\frac{96056179}{180000}\Bigr)
\nn \\ &-
	C_F^2 C_A\Bigl(64 \zeta_3^2+22 \zeta_6-232 \zeta_2 \zeta_3-8 \zeta_5+\frac{19883}{18}\zeta_4-\frac{74378}{45} \zeta_3
	+\frac{392641 }{216}\zeta_2-\frac{113349701}{51840}\Bigr)
\nn \\ &+
	C_F C_A n_f \Bigl(-72 \zeta_3 \zeta_2-\frac{200}{3}\zeta_5+\frac{812}{9}\zeta_4+\frac{61169}{270}\zeta_3-\frac{3933857}{16200} \zeta_2+\frac{167350393}{2160000}\Bigr)
\nn \\ &+
	C_F^2 n_f \Bigl(\frac{224}{3} \zeta_3 \zeta_2+\frac{128}{3}\zeta_5-\frac{34}{9}\zeta_4-\frac{7726}{45}\zeta_3+\frac{98803}{360}\zeta_2-\frac{406426043}{972000}\Bigr)
\nn \\ &+
	\frac{d_{abc}d^{abc}}{N_R} N_{F,V} \Bigl(\frac{40}{3}\zeta_5+\frac{\zeta_4}{2}-\frac{19}{3}\zeta_3-5 \zeta_2-\frac{1}{6}\Bigr)
\,,\end{align}
where we adopt the same convention for the $d_{abc}d^{abc}$ color structure as in \eq{EEC_z1_N3LO_ee}.

We stress that to calculate the \moment using \eq{sumRulez1} we need the $\cO(\alpha_s^3)$ logarithmic coefficients also in the forward ($z\to 0$) limit, whose factorization theorem was obtained in \refcite{Dixon:2019uzg}.
However, at the time of the publication of \refcite{Dixon:2019uzg}, the NNLO time-like splitting function $P_{qg}^{(2)}$ available in the literature was not correct.
A new result for $P_{qg}^{(2)}$ was more recently obtained in \refcite{Chen:2020uvt} and we have confirmed this result in our calculation in \refcite{Ebert:2020qef}.
Since $P_{qg}^{(2)}$ is part of the singlet time-like splitting kernel matrix governing the logarithmic structure of the EEC in the $z\to 0$ limit~\cite{Dixon:2019uzg}, the new result for $P_{qg}^{(2)}$ modifies the small angle limit of the Higgs EEC at $\cO(\alpha_s^3)$ and beyond.
The result in $e^+e^-$ is not modified since in that case $P_{qg}^{(2)}$ contributes to the $z\to0$ logarithmic coefficients only starting at $\cO(\alpha_s^4)$.
In the Higgs case, using the correct splitting function  we obtain
{
\allowdisplaybreaks
\begin{align}\label{eq:sumrule_H}
\frac{1}{\born} &\int_0^1 \df z \, z \EECregaslab{3}{H}
 \nn\\&
 =  C_A^3 \Bigl(56  \zeta_3^2+591  \zeta_6-132  \zeta_2 \zeta_3-356  \zeta_5-\frac{1345}{3}  \zeta_4-\frac{259187}{225}  \zeta_3
	- \frac{1722547  \zeta_2}{2250}+\frac{168476576111}{24300000}\Bigr)
\nn\\&+
	C_A^2 n_f \Bigl(-\frac{166}{3}  \zeta_2 \zeta_3+92  \zeta_5-\frac{4705}{36}  \zeta_4+\frac{182159}{450}  \zeta_3+\frac{8938517}{13500}\zeta_2-\frac{194619951647}{48600000}\Bigr)
\nn\\&+
	C_A C_F n_f \Bigl(\frac{208}{3}  \zeta_2 \zeta_3-120  \zeta_5-\frac{263}{15}
    \zeta_4+\frac{48022}{225}  \zeta_3-\frac{48299}{540}\zeta_2-\frac{3671238613}{9720000}\Bigr)
\nn\\&+
   C_F^2 n_f \Bigl(80  \zeta_5+\frac{308}{15}  \zeta_4-\frac{19342}{225}  \zeta_3+\frac{1333}{100}  \zeta_2-\frac{1101193}{97200}\Bigr)
\nn\\&+
   C_A n_f^2 \Bigl(\frac{284}{9}\zeta_4-\frac{1234}{45}  \zeta_3-\frac{34771}{225}  \zeta_2+\frac{427251479}{607500}\Bigr)
\nn\\&+
   C_F n_f^2 \Bigl(\frac{8}{9}
   \zeta_4-\frac{536}{15}  \zeta_3-\frac{6967}{900}  \zeta_2+\frac{173703509}{1620000}\Bigr)
\nn\\&+
n_f^3 \Bigl(\frac{596}{45}\zeta_2-\frac{4947899}{121500}\Bigr)\,.
\end{align}
}
The \moment of the EEC for $e^+e^-$ and Higgs at $\cO(\alpha_s^3)$ in \eqs{sumrule_ee}{sumrule_H} constitute two new results.
In particular it is the first piece of information on the EEC in QCD analytically at NNLO in the bulk of the distribution.
As a matter of fact, at NNLO the EEC is only known numerically in QCD for $e^+e^-$ \cite{DelDuca:2016csb,Tulipant:2017ybb} while no results at this order at all are available for the EEC in gluon-initiated Higgs decay.
Noting that the full result for the EEC at this order is known in $\cN=4$~\cite{Henn:2019gkr}, we can check if the $\cN=4$ result indeed constitutes the leading transcendental terms of the QCD result.
While the result of \refcite{Henn:2019gkr} is expressed in terms of a two-fold integral and therefore cannot be directly checked analytically, in \refcite{Korchemsky:2019nzm} the \moment of the distribution away from the endpoints has been obtained and reads
\begin{align}
\frac{1}{\born} \int_0^1 \df z \,z\, \frac{\df \sigma_{\text{reg}}^{\cN=4}}{\df z}
 = 2 \Bigl(\frac{N_c \alpha_s}{\pi}\Bigr)^3
   \Bigl(-2 +\frac{2}{3}\pi^2-\frac{11}{8}\zeta_3+\frac{\pi^4}{80}-\frac{\pi^2 \zeta_3}{12}-\frac{5}{4}\zeta_5+\frac{197}{40320}\pi^6+\frac{7}{16}\zeta_3^2 \Bigr)
\,.\end{align}
We see that this result has no uniform transcendentality, but the leading transcendental piece exactly matches our result in QCD after taking $C_A, C_F \to N_c$. Therefore we confirm that the principle of maximal transcendentality holds also at NNLO for the \moment of the distribution in the bulk.

\section{\boldmath Resummation of the EEC at \texorpdfstring{N$^3$LL$^\prime$}{N3LL'} accuracy}
\label{sec:resummation}

In this section, we use our results to obtain for the first time the EEC spectrum in the back-to-back limit resummed at N$^3$LL$^\prime$ accuracy.
In \sec{EEC_RGE}, we review how the large logarithms $\ln(1-z)$ can be resummed to all-orders by solving the renormalization group equations (RGEs) of the hard, jet and soft functions.
Details of its implementation are presented in \sec{EEC_numerics_setup}, before we show our numeric results in \sec{EEC_resummed_results}.

\subsection{Renormalization group evolution}
\label{sec:EEC_RGE}

The hard, jet and soft functions entering \eqs{EEC_fact_thm_q}{EEC_fact_thm_g} obey
the same RGEs as the hard, beam and soft functions in TMD factorization,
\begin{alignat}{3} \label{eq:RGEs}
 \frac{\df}{\df \ln \mu} \ln H_i(Q,\mu) &
 = \gamma_H^i(Q,\mu)
\,,\nn\\
 \frac{\df}{\df \ln \mu} \ln J_i(b_T,\mu,\nu/Q) &
 = \tilde\gamma_J^i(\mu,\nu/Q)
\,,\qquad&&
 \frac{\df}{\df \ln \nu} \ln J_i(b_T,\mu,\nu/Q) &&
 = -\frac12 \tilde\gamma_\nu^i(b_T,\mu)
\,,\nn\\
 \frac{\df}{\df \ln \mu} \ln \tilde S_i(b_T,\mu,\nu) &
 = \tilde\gamma_S^i(\mu,\nu)
\,,\qquad&&
 \frac{\df}{\df \ln \nu} \ln \tilde S_i(b_T,\mu,\nu) &&
 = \tilde\gamma_\nu^i(b_T,\mu)
\,.\end{alignat}
The anomalous dimensions with respect to $\mu$ have the all-order form
\begin{align} \label{eq:mu_anom_dims}
 \gamma_H^i(Q,\mu) &= 4 \GammaC^i[\as(\mu)] \ln\frac{Q}{\mu} + 4 \gamma_i[\as(\mu)]
\,,\nn\\
 \tilde\gamma_J^i(\mu,\nu/Q) &= 2 \GammaC^i[\as(\mu)] \ln\frac{\nu}{Q} + \tilde\gamma_J^i[\as(\mu)]
\,,\nn\\
 \tilde\gamma_S^i(\mu,\nu) &= 4 \GammaC^i[\as(\mu)] \ln\frac{\mu}{\nu} + \tilde\gamma_S^i[\as(\mu)]
\,.\end{align}
Here, $\GammaC^i(\as)$ is the cusp anomalous dimension, $\gamma_q(\as)$ and $\gamma_g(\as)$ are the quark and gluon anomalous dimensions,
and $\tilde\gamma_J^i(\as)$ and $\tilde \gamma_S^i(\as)$ are the jet and soft noncusp anomalous dimensions, respectively.
In \eq{mu_anom_dims}, the notation $\gamma[\as(\mu)]$ indicates that their scale dependence only arises through the strong coupling constant $\as(\mu)$,
and we distinguish the noncusp anomalous dimensions from the full anomalous dimension by the number of arguments.
Note that the jet anomalous dimension is identical to that of the TMD beam and fragmentation functions,
and in particular $\tilde\gamma_J^i(\as) \equiv \tilde\gamma_B^i(\as)$.
$\mu$ independence of the cross section implies that
\begin{align} \label{eq:sum_mu_anom_dims}
 \gamma_H^i(Q,\mu) + 2 \tilde\gamma_J^i(\mu,\nu/Q) + \tilde\gamma_S^i(\mu,\nu) = 0
\,.\end{align}
The rapidity anomalous dimension itself obeys an RGE,
\begin{align} \label{eq:gamma_nu_RGE}
 \frac{\df}{\df\ln\mu} \tilde\gamma_\nu^i(b_T,\mu) = -4 \GammaC^i[\as(\mu)]
\,,\end{align}
which follows by commutativity of applying both $\df/\df\ln\mu$ and $\df/\df\ln\nu$ to either $J_i$ or $\tilde S_i$.
The solution to \eq{gamma_nu_RGE} can be written as
\begin{align} \label{eq:nu_anom_dim}
 \tilde\gamma_\nu^i(b_T,\mu) &
 = -4 \int_{\mu_0}^\mu \frac{\df\mu'}{\mu'} \GammaC^i[\as(\mu')] + \tilde\gamma_{\nu,{\rm FO}}^i(b_T, \mu_0)
\,,\end{align}
where the integral resums logarithms $\ln(\mu/\mu_0)$, leaving only logarithms $\ln(b_T \mu_0 / b_0)$
in the boundary term that is evaluated at fixed order, as indicated.
This generalizes \eq{jet_anom_dims} to arbitrary boundary scales $\mu_0$.
\Eq{jet_anom_dims} is obtained by choosing the canonical scale $\mu_0 = b_0/b_T$,
which eliminates all large logarithms in the boundary term such that it can be reliably calculated in fixed order.
The boundary term at this particular choice is commonly abbreviated as
\begin{align}
 \tilde\gamma_\nu^i[\as(b_0/b_T)] = \tilde\gamma_{\nu,{\rm FO}}^i(b_T,\mu = b_0/b_T)
\,.\end{align}
As for the $\mu$ anomalous dimensions, the boundary term is distinguished
from the full $\tilde\gamma_\nu^i(b_T,\mu)$ by the number of arguments.

We note that the rapidity anomalous dimension becomes nonperturbative for $b_T \gtrsim \lqcd^{-1}$,
irrespective of whether $\mu$ is perturbative or not. In our numeric analysis, we will simply freeze out $\mu_0$
to avoid the Landau pole, but note that it would be very interesting to extract nonperturbative contributions
to $\tilde\gamma_\nu$ using EEC data from LEP. Another promising approach is to calculate it using lattice QCD
as suggested recently~\cite{Ebert:2018gzl,Ebert:2019okf,Vladimirov:2020ofp},
and first exploratory results have recently been obtained in \refscite{Shanahan:2020zxr,Zhang:2020dbb},
see also \refcite{Vladimirov:2020umg} for first estimates of its large-$b_T$ asymptotics.

By solving \eq{RGEs}, one can evolve the hard, jet and soft functions from their natural scales to the common scales $(\mu,\nu)$.
This two-dimensional evolution is independent of the chosen path by virtue of \eq{nu_anom_dim},
and we choose to first evolve in virtuality $\mu$ before evolving in rapidity $\nu$.
The resummed cross section in \eq{EEC_fact_thm_q} is then given by
\begin{align} \label{eq:EEC_resummed}
 \frac{\df\sigma}{\df z} &
 = \frac{\born}{8} \int_0^\infty \!\!\df (b_T Q)^2 \, J_0\bigl(b_T Q \sqrt{1-z}\bigr)
     H_{q\bq}(Q,\mu_H) J_q\Bigl(b_T, \mu_J, \frac{\nu_J}{Q}\Bigr) J_\bq\Bigl(b_T, \mu_J, \frac{\nu_J}{Q}\Bigr) \tilde S_q(b_T, \mu_S, \nu_S)
 \nn\\&\quad \times
 \exp\left[ \int_{\mu_H}^\mu \frac{\df\mu'}{\mu'} \gamma_H^q(Q, \mu')
          + 2 \int_{\mu_J}^\mu \frac{\df\mu'}{\mu'} \tilde\gamma_J^q(\mu', \nu_J/Q)
          + \int_{\mu_S}^\mu \frac{\df\mu'}{\mu'} \tilde\gamma_S^q(\mu', \nu_S) \right]
 \Bigl(\frac{\nu_J}{\nu_S}\Bigr)^{\tilde\gamma_\nu^q(b_T, \mu)}
 \nn\\&\quad \times
 \bigl[ 1 + \cO(1-z) \bigr]
\,,\end{align}
and similar for the gluon case shown in \eq{EEC_fact_thm_g}.
\Eq{EEC_resummed} is manifestly independent of the overall rapidity scale $\nu$,
while the dependence on $\mu$ cancels due to \eq{sum_mu_anom_dims}.%
\footnote{The integrals over $\mu'$ are often implemented using an approximate analytic solution
which leads to a small residual dependence on $\mu$~\cite{Bell:2018gce,Billis:2019evv}.
To avoid this effect, we have implemented all integrals and the running coupling constant exactly,
but we have checked that the difference to the analytic approximation is negligible.}
By choosing the canonical resummation scales as
\begin{align} \label{eq:canonical_scales}
 \mu_H \sim Q \,,\quad &\mu_J \sim \frac{b_0}{b_T} \,, \quad \mu_S \sim \frac{b_0}{b_T}
\,,\nn\\
 &\nu_J \sim Q \,,\quad~~ \nu_S \sim \frac{b_0}{b_T}
\,,\end{align}
the fixed-order boundary terms in the first line of \eq{EEC_resummed} are free of large logarithms and can be reliably evaluated in fixed-order perturbation theory, while all large logarithms are explicitly exponentiated.

The logarithmic accuracy of the resummed cross section in \eq{EEC_resummed} is classified by the exponentiated logarithms.
For example, LL refers to exponentiating all terms $\as L^2$, requiring only the one-loop cusp anomalous dimension and beta function.
The fixed-order boundary terms $H, J$ and $S$ must be chosen at an order such that they cancel all logarithms obtained from expanding the exponential in fixed order.
In practice, one often chooses the boundary terms at one order higher as this significantly reduces residual scale dependencies,
which is referred to as prime counting~\cite{Almeida:2014uva}.
For completeness, \tbl{log_counting} lists the ingredients required up to N$^3$LL$^\prime$.

{
\renewcommand{\arraystretch}{1.2}
\begin{table}[pt]
\centering
 \begin{tabular}{l|c|c|c|c} \hline\hline
  Accuracy & $H$, $J$, $S$ & $\GammaC(\as)$ & $\gamma(\as)$ & $\beta(\as)$ \\\hline
  LL           & Tree level & $1$-loop & --       & $1$-loop \\\hline
  NLL          & Tree level & $2$-loop & $1$-loop & $2$-loop \\\hline
  NLL$^\prime$ & $1$-loop   & $2$-loop & $1$-loop & $2$-loop \\\hline
  NNLL         & $1$-loop   & $3$-loop & $2$-loop & $3$-loop \\\hline
  NNLL$^\prime$& $2$-loop   & $3$-loop & $2$-loop & $3$-loop \\\hline
  N$^3$LL         & $2$-loop   & $4$-loop & $3$-loop & $4$-loop \\\hline
  N$^3$LL$^\prime$& $3$-loop   & $4$-loop & $3$-loop & $4$-loop \\\hline
 \hline
 \end{tabular}
\caption{%
Classification of the resummation accuracy in terms of the
fixed-order expansions of boundary term, anomalous dimensions and beta function.
}
\label{tbl:log_counting}
\end{table}
\renewcommand{\arraystretch}{1.0}
}

At N$^3$LL$^\prime$, we need to implement the hard, jet and soft function at three loops.
The Drell-Yan hard function is known at $\cO(\as^3)$ from the quark form factor%
~\cite{Kramer:1986sg, Matsuura:1987wt, Matsuura:1988sm, Gehrmann:2005pd, Moch:2005tm, Moch:2005id, Baikov:2009bg, Lee:2010cga, Gehrmann:2010ue}, and is explicitly given in \refcite{Gehrmann:2010ue}.%
\footnote{Note that starting at three loops, the exchanged vector boson can couple to closed quark loops,
whose couplings differ from those of the Born process.
This is the origin of the $N_{F,V}$ piece in \refcite{Gehrmann:2010ue}.
For our numerical illustration later, we simply set $N_{F,V} = 0$.}
The soft function has been calculated at $\cO(\as^3)$ in \refcite{Li:2016ctv} and confirmed in ref.~\cite{Ebert:2020yqt},
while the EEC jet function at three loops is the remaining ingredient provided in this paper.
The resummation also requires knowledge of the four-loop cusp anomalous dimension~\cite{Korchemsky:1987wg, Moch:2004pa, Vogt:2004mw, Lee:2016ixa,Moch:2017uml,Lee:2019zop,Henn:2019rmi,Bruser:2019auj, Henn:2019swt, vonManteuffel:2020vjv}
(see \refcite{Bruser:2019auj} for a complete list of partial four-loop results).
The three-loop quark and gluon anomalous dimensions $\gamma_{q,g}$ are known from the corresponding form factors~\cite{Kramer:1986sg, Matsuura:1987wt, Matsuura:1988sm, Harlander:2000mg, Gehrmann:2005pd, Moch:2005id, Moch:2005tm} (see \refcite{vonManteuffel:2020vjv} for the result at four loops).
The beam and soft noncusp anomalous dimensions $\tilde\gamma_{B,S}^i$ were first obtained by consistency with the invariant-mass dependent jet function and threshold soft function, all of which are known at three loops~\cite{Vogt:2004mw,Moch:2004pa,Stewart:2010qs,Berger:2010xi,Bruser:2018rad,Banerjee:2018ozf}, and were confirmed by explicit calculations in \refscite{Li:2016ctv, Luo:2019szz,Ebert:2020yqt}.
The rapidity anomalous dimension $\tilde\gamma_\nu(\as)$ is known at N$^3$LO~\cite{Luebbert:2016itl, Li:2016ctv, Vladimirov:2016dll}.
Finally, resummation at N$^3$LL accuracy also requires knowledge of the QCD $\beta$ function at four loops~\cite{Tarasov:1980au, Larin:1993tp, vanRitbergen:1997va, Czakon:2004bu}.
Explicit expressions for all these anomalous dimension through N$^3$LO in the notation used in this paper are collected in \refscite{Ebert:2017uel, Billis:2019vxg}.

\subsection{Resummation scales and perturbative uncertainties}
\label{sec:EEC_numerics_setup}

We choose the canonical resummation scales as
\begin{align} \label{eq:scales}
 \mu_H &= \nu_J = Q
\,,\quad
 \mu_J = \mu_S = \mu_0 = \frac{b_0}{b_T^*(b_T)}
\,,\quad
 \nu_S = \frac{b_0}{b_T}
\,.\end{align}
Here, we employ a local $b^*$ prescription to freeze out the virtuality scales to avoid the Landau pole at large $b_T$, with
\begin{align}
 b_T^*(b_T) = \frac{b_T}{\sqrt{1 + b_T^2/b_{\rm max}^2}}
\,,\qquad
 \frac{b_0}{b_{\rm max}} = 1~\GeV
\,.\end{align}
The functional form in \eq{scales} is identical to the $b^*$ prescription of \refscite{Collins:1981uk,Collins:1981va},
but following \refcite{Lustermans:2019plv} we only modify the resummation scales rather than globally replacing $b_T$ by $b_T^*(b_T)$,
as this would induce a global power correction $\cO(b_T^2/b_{\rm max}^2)$.

Note that we always choose (variations around) the canonical  resummation scales in \eq{scales}.
In a detailed phenomenological study, one would smoothly turn off the resummation when the power corrections
to \eq{EEC_resummed} become comparable to the terms predicted by the factorization theorem.
Here, we refrain from doing so, as we only intend to illustrate the impact of the new three loop results on the resummation
in the regime where canonical resummation is justified.

To estimate perturbative uncertainties, we follow the procedure developed in \refcite{Stewart:2013faa}
and separately consider a fixed-order and a resummation uncertainty, and in addition consider an uncertainty from our nonperturbative prescription.
Since these sources are considered uncorrelated, the individual uncertainties are then added in quadrature,
\begin{align} \label{eq:delta}
 \Delta_{\rm tot} = \sqrt{\Delta_{\rm fo}^2 + \Delta_{\rm res}^2 + \Delta_{\rm np}^2 }
\,,\end{align}
which we apply symmetrically around the central prediction.

The fixed-order uncertainty is estimated by varying all scales except $\mu_0$ in \eq{scales}
by a common factor of $1/2$ or $2$ and take $\Delta_{\rm fo}$ to be the maximum deviation.
This probes all fixed-order boundary terms in the first line of \eq{EEC_resummed},
but does not affect the exponentiated logarithms in the second line,
and thus is akin to a standard fixed-order scale variation.

The resummation uncertainty is probed by individually varying all scales by a factor of $1/2$ or $2$ around their central value,
constrained such that the arguments of all exponentiated logarithms in the second line of \eq{EEC_resummed}
are varied up or down by a factor of at most $2$. We do not vary $\mu_H$, whose variation is already covered by
the fixed-order uncertainty, resulting in 35 variations in total~\cite{Stewart:2013faa}.
These variations probe the cancellation of the exponentiated logarithms with those in the fixed-order boundary conditions,
and thus are interpreted as a resummation uncertainty. Since these variations can be highly correlated,
we define $\Delta_{\rm res}$ as the envelope of all variations.

Finally, we vary $b_0 / b_{\rm max} = 0.5, 2~\GeV$ to probe the uncertainty $\Delta_{\rm np}$ of our nonperturbative prescription.

\subsection{Numerical results}
\label{sec:EEC_resummed_results}

We illustrate the impact of our new results by numerically studying the EEC spectrum differential in the angle $\chi$,
\begin{align}
 \frac{\df\sigma}{\df\chi} = \frac12 \sin\chi \frac{\df\sigma}{\df z} \bigg|_{z = \frac12(1-\cos\chi)}
\,,\end{align}
for photon-induced hadron production up to N$^3$LL$^\prime$.
We always work on the $Z$-pole with $Q = m_Z = 91.1876~\GeV$, and choose $\as(m_Z) = 0.118$ and evolve it according to \tbl{log_counting}.
The resummed spectra are evaluated using an implementation in \texttt{SCETlib}~\cite{scetlib}.

\begin{figure*}
 \centering
 \includegraphics[width=0.48\textwidth]{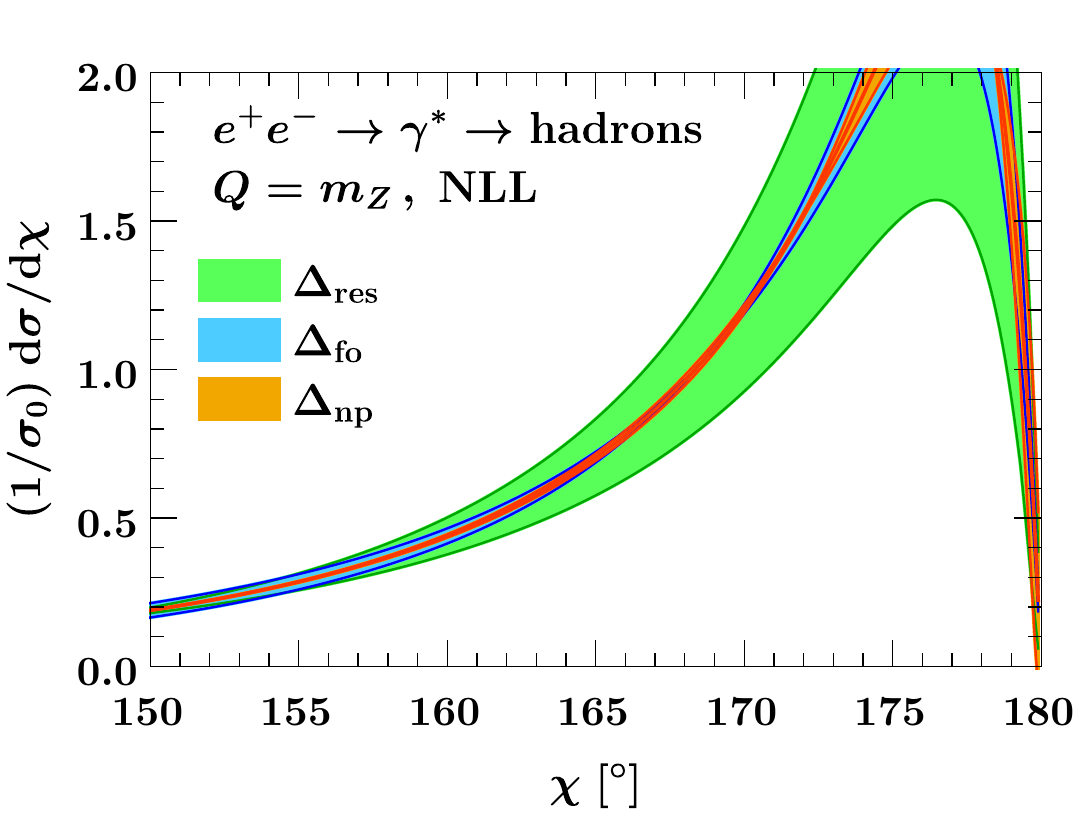}
 \hfill
 \includegraphics[width=0.48\textwidth]{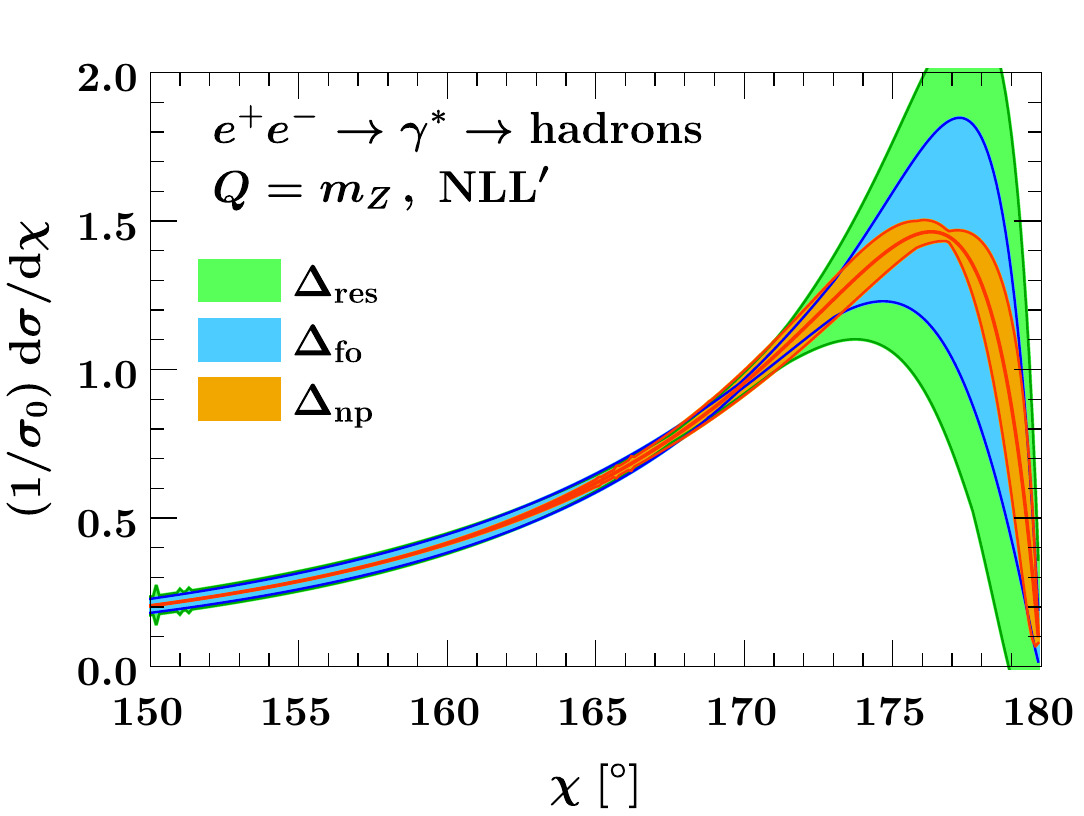}
 \\
 \includegraphics[width=0.48\textwidth]{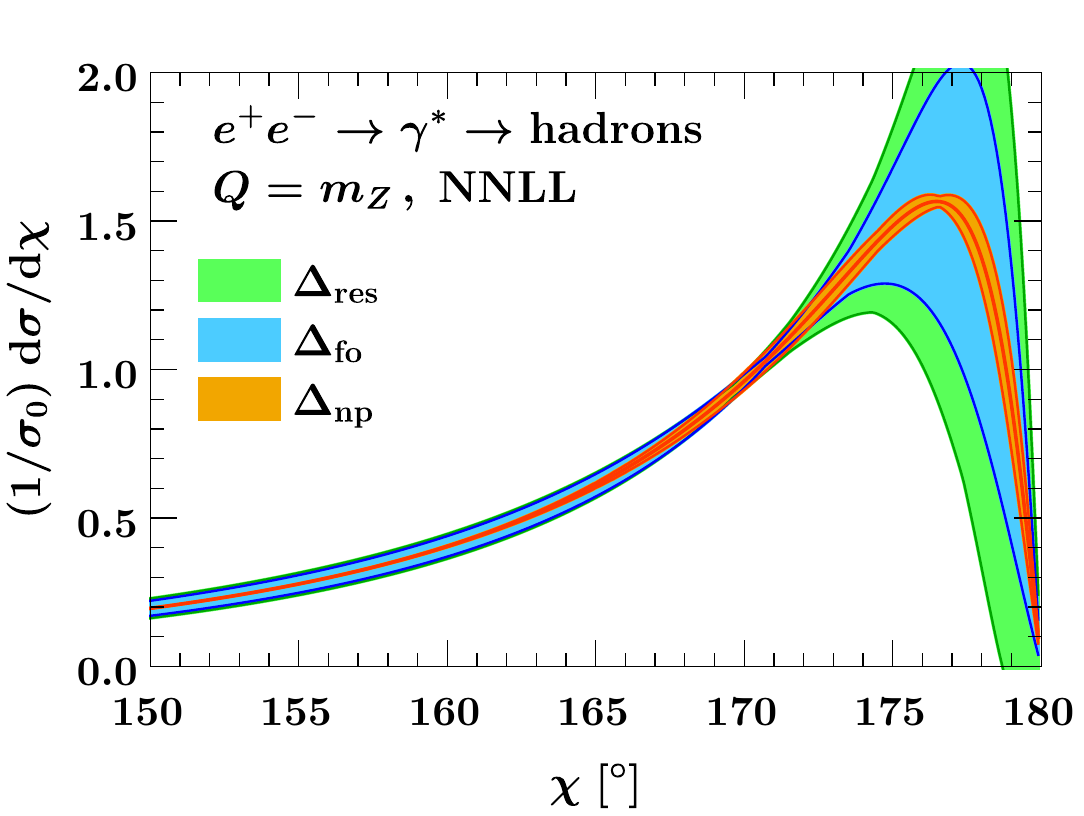}
 \hfill
 \includegraphics[width=0.48\textwidth]{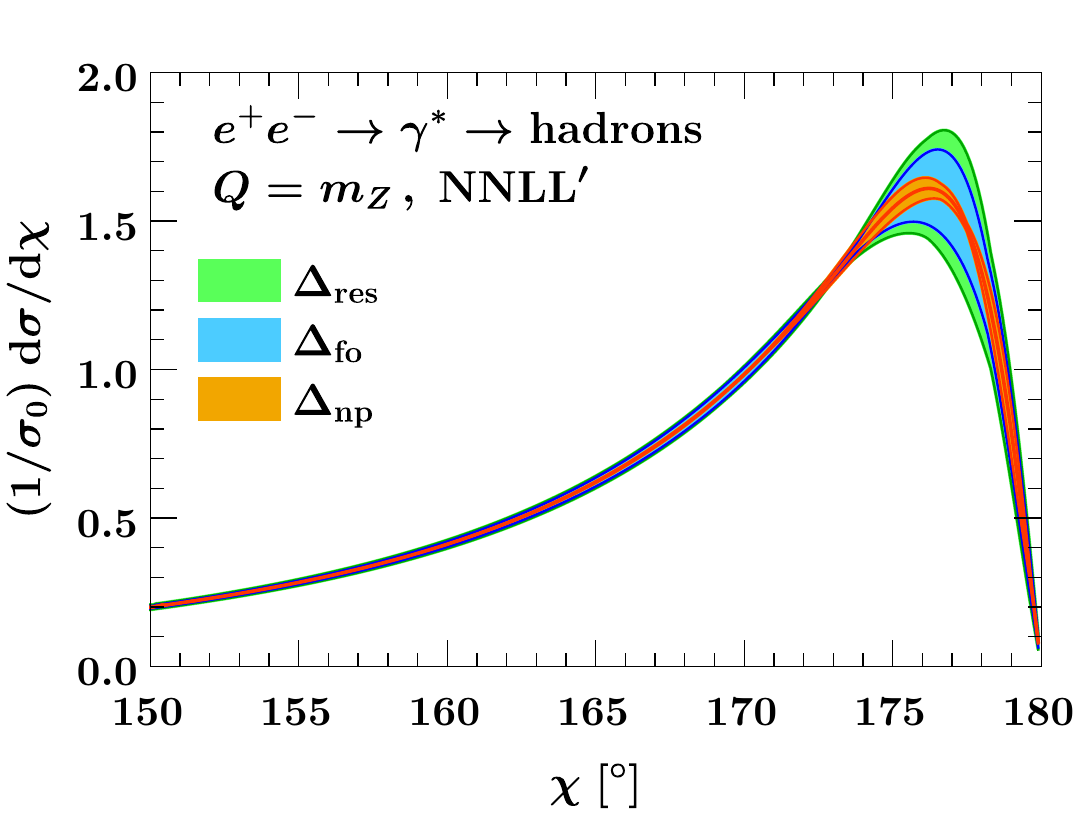}
 \\
 \includegraphics[width=0.48\textwidth]{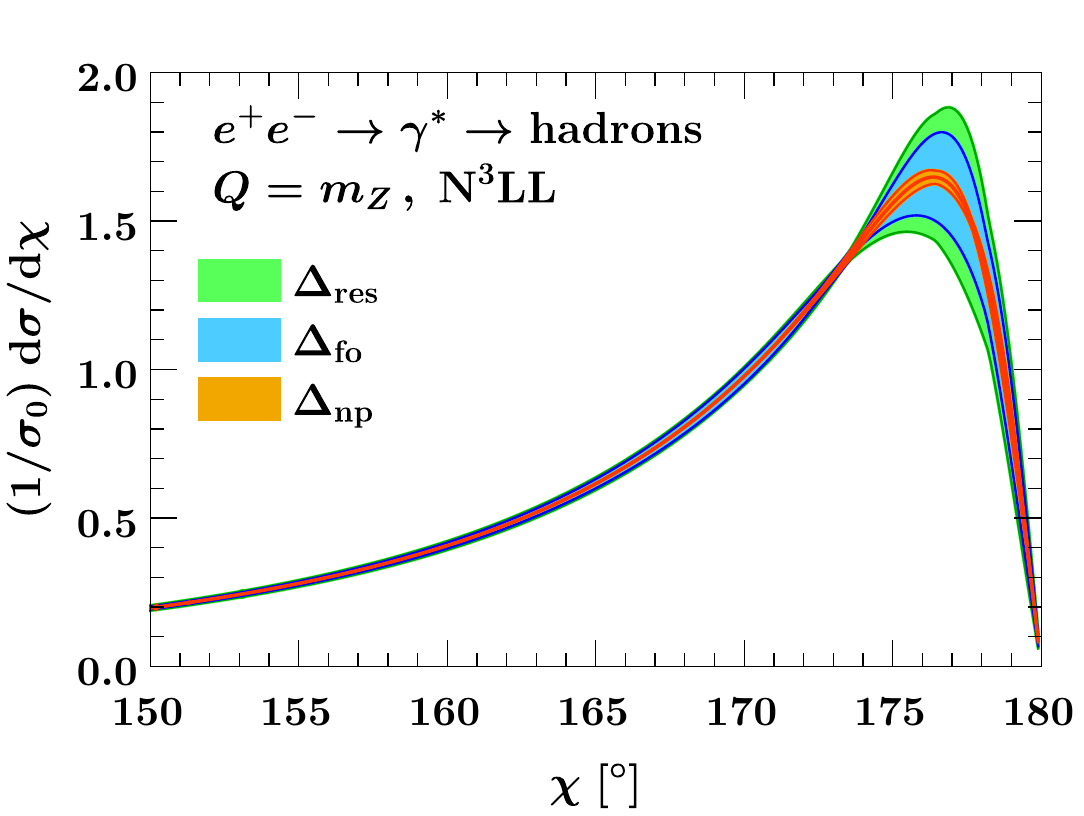}
 \hfill
 \includegraphics[width=0.48\textwidth]{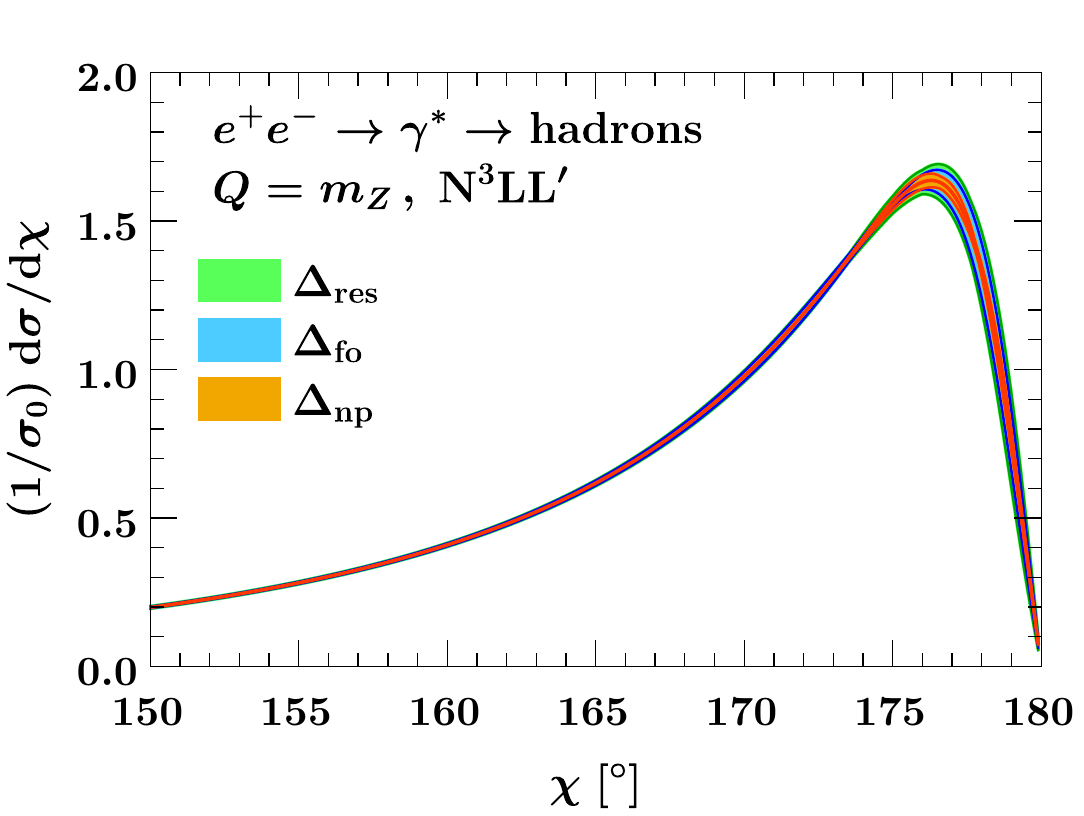}
 \caption{Breakdown of the total uncertainty into fixed-order ($\Delta_{\rm fo}$, blue), resummation ($\Delta_{\rm res}$, green) and nonperturbative ($\Delta_{\rm np}$) uncertainties for different resummation orders.}
 \label{fig:EEC_uncertainties}
\end{figure*}

Figure~\ref{fig:EEC_uncertainties} ~shows the breakdown of the total uncertainty into the different sources,
namely fixed-order ($\Delta_{\rm fo}$, blue), resummation ($\Delta_{\rm res}$, green) and nonperturbative ($\Delta_{\rm np}$, orange) uncertainties,
for all resummation orders except LL.
In all cases, the smallest uncertainty is $\Delta_{\rm np}$ from varying $b_{\rm max}$, and only becomes relevant for the very precise N$^3$LL$^\prime$ prediction.
The resummation uncertainty $\Delta_{\rm res}$ is always larger than $\Delta_{\rm fo}$, but both are of comparable size at NNLL and higher.
As is clear from \fig{EEC_uncertainties}, the overall uncertainties reduce drastically with increasing the resummation accuracy, i.e.~as NLL $\to$ NNLL $\to$ N$^3$LL.

A striking feature is the huge reduction of uncertainties when going from N$^n$LL (left panel) to N$^n$LL$^\prime$ (right panel),
which has a much bigger impact than simply increasing the resummation order from N$^{n}$LL to N$^{n+1}$LL.
This illustrates the importance of prime counting, i.e.~including the fixed-order boundary terms in the resummation
at the same perturbative order as the resummation (confer \tbl{log_counting}).

We remark that we encounter much larger uncertainties than observed in the NNLL results in \refscite{deFlorian:2004mp,Tulipant:2017ybb,Kardos:2018kqj}.
Our uncertainties  are dominated by variations of the low scales, as they probe $\as(1/b_T^*)$ and thus become large at large $b_T$.
Due to applying these uncertainties symmetryically, the resulting uncertainty bands even become negative up to NNLL,
but greatly improved beyond NNLL.
We will address this in more detail in \sec{literature}.

\begin{figure*}
 \centering
 \includegraphics[width=0.48\textwidth]{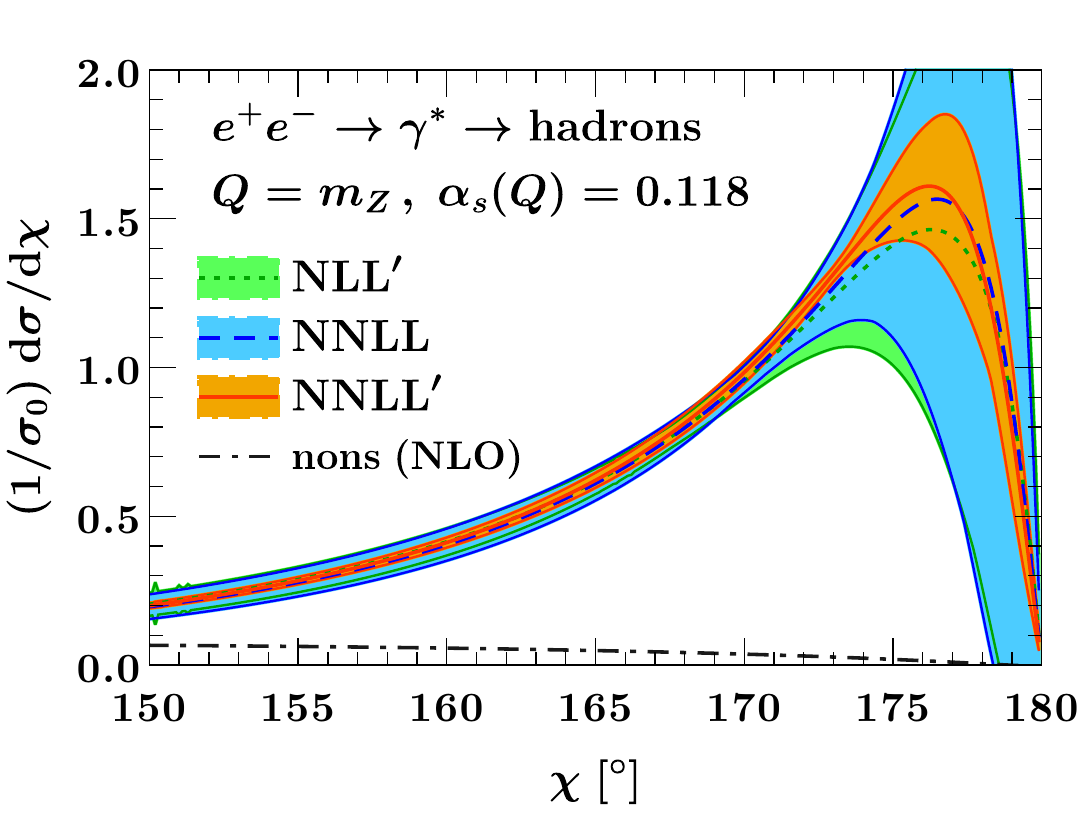}
 \hfill
 \includegraphics[width=0.48\textwidth]{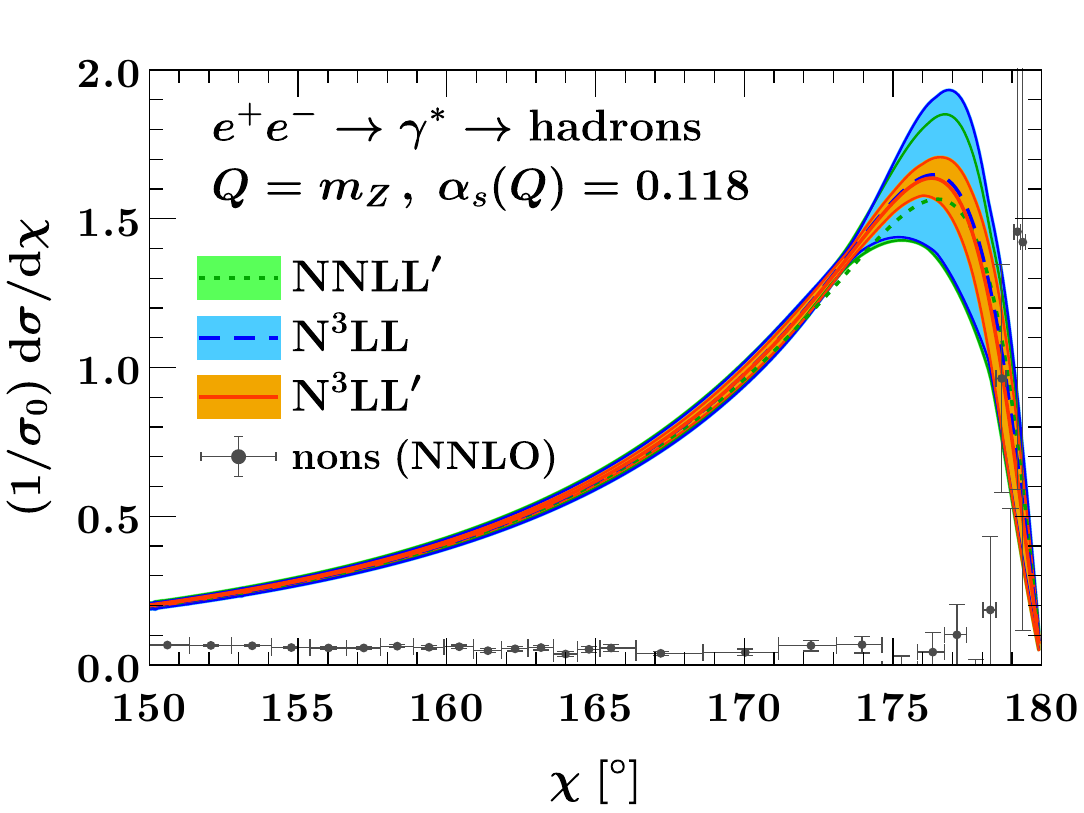}
 \caption{Comparison of the resummed EEC spectrum as a function of the angle $\chi$ at different resummation orders.
          The shaded areas, bounded by the solid lines, show the individual uncertainties, with the central value indicated according to the legend.}
 \label{fig:EEC_resummed}
\end{figure*}

Figure~\ref{fig:EEC_resummed} shows a comparison of the resummed EEC spectrum at various orders.
In the left panel, we compare NLL$^\prime$ through NNLL$^\prime$, while in the right panel we compare NNLL$^\prime$ through N$^3$LL$^\prime$.
In both cases we also plot the nonsingular distribution defined as
\beq
	 \frac{\df\sigma^\text{nons}}{\df z} \equiv \frac{\df\sigma}{\df z} - \frac{\df\sigma_1}{\df z}
\,,\eeq
where $\df\sigma_1/\df z$ is the leading-power cross section in the back-to-back limit as predicted by the factorization formula in \eq{EEC_fact_thm_q} and presented explicitly in \sec{EEC_back_to_back} up to $\cO(\alpha_s^3)$.
Thus, $\df\sigma^\text{nons}/\df z$ shows the impact of the fixed-order corrections beyond leading power.
In the left plot, the dot-dashed black line shows the neglected correction from the NLO fixed-order results calculated in \refcite{Dixon:2018qgp} (with respect to the $\cO(\alpha_s^2)$ leading power result), while in the right plot we show the nonsingular distribution at $\cO(\alpha_s^3)$ using the NNLO fixed-order results calculated numerically in~\refcite{Tulipant:2017ybb}.%
\footnote{We thank G\'abor Somogyi for private communication of these results}
As expected, these fixed-order corrections are neglibile for  $\chi \gtrsim 160^\circ$, thus justifying employing canonical leading power resummation in the shown range.
One notable feature in \fig{EEC_resummed} is that going from NLL$^\prime$ to NNLL actually increases the size of the scale variations, except in the region $\chi \sim 170-177^\circ$.
Nevertheless, the central value at NNNL (blue dashed) is much closer to the central value at NNLL$^\prime$ (red solid) than at NLL (green dotted).
We observe a similar pattern when comparing NNLL$^\prime$ and N$^3$LL in the right panel, even though the effects are less pronounced at this order.
Overall, the central values show very good convergence beyond NNLL, with greatly reduced uncertainties at N$^3$LL$^\prime$ of about $\pm 4\%$ at the peak, compared to about $\pm15\%$ at N$^3$LL.

\begin{figure}
 \centering
 \includegraphics[width=0.7\textwidth]{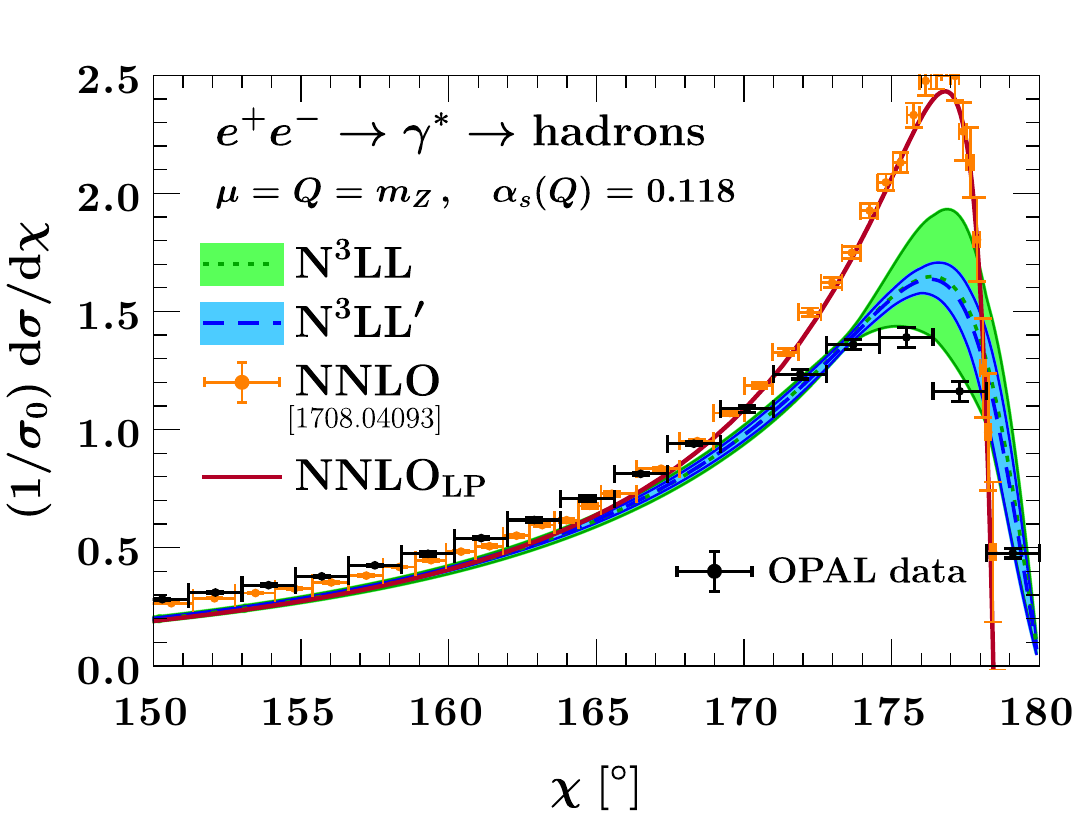}
 \caption[dsadsasd]{Comparison of the N$^3$LL and N$^3$LL$^\prime$ resummed EEC partonic spectrum with the fixed order NNLO numerical calculation of~\refcite{Tulipant:2017ybb}, the LEP data from the OPAL collaboration \cite{Acton:1993zh}, and the analytic leading power spectrum at $\cO(\alpha_s^3)$\textsuperscript{\ref{footnote1}} from \eq{EEC_z1_N3LO_ee}.}
 \label{fig:EEC_data_comp}
\end{figure}

In \fig{EEC_data_comp}, we overlay the N$^3$LL and N$^3$LL$^\prime$ resummed spectra of \fig{EEC_resummed} with the fixed order NNLO numerical calculation of~\refcite{Tulipant:2017ybb} as well as the experimental measurement of the EEC \cite{Acton:1993zh} from the OPAL collaboration at LEP.
We observe that resummation effects improve substantially the behavior of the spectrum in the peak region compared to the fixed order calculation. We also note that pushing the resummation to N$^3$LL$^\prime$ accuracy is crucial to achieve a perturbative control on the theory prediction that is competitive with the experimental uncertainty on this observable.
However, it is important to point out that for a realistic comparison with experimental data and for the extraction of the strong coupling constant from this event shape, it is essential to include hadronization effects~\cite{Abreu:1990us,Acton:1991cu,Acton:1993zh,Abreu:1993kj,Abe:1994mf,Korchemsky:1999kt,Tulipant:2017ybb,Kardos:2018kqj,dEnterria:2019its}, as resummation effects alone, even at  N$^3$LL$^\prime$, cannot fully bridge the gap between partonic fixed order results and data. For completeness, we also show in \fig{EEC_data_comp}  the analytic leading power spectrum in the back-to-back limit at $\cO(\alpha_s^3)$%
\footnote{Note that, as explained in the Introduction, in the context of factorization theorems it is customary to count $\cO(\alpha_s^n)$ contributions as N$^n$LO. Therefore, the leading power spectrum for $z \to 1$ at $\cO(\alpha_s^3)$ is referred as the EEC in the back-to-back limit at N$^3$LO, hence the title of \sec{EEC_back_to_back}. However, since in \fig{EEC_data_comp} we included the leading power fixed order result mainly for comparison with the numerical calculation of~\refcite{Tulipant:2017ybb}, which counts $\cO(\alpha_s^{n+1})$ corrections as N$^n$LO as it is usually done for fixed order calculations of event shapes, we refrained from mixing the notation and therefore labeled it as $\text{NNLO}_\text{LP}$ in the legend.\label{footnote1}}  
(in red) from \eqs{EEC_z1_NLO_NNLO}{EEC_z1_N3LO_ee}.
In the peak region, this result perfectly agrees with the fixed order NNLO numerical calculation of~\refcite{Tulipant:2017ybb},
thereby again justifying resummation in this region. Note that the difference between those two results is exactly the non-singular distribution plotted in the right panel of \fig{EEC_resummed}.
\subsection{Comparison to literature}
\label{sec:literature}

Resummed predictions for the EEC in the back-to-back limit have previously been reported in
\refscite{deFlorian:2004mp,Tulipant:2017ybb,Kardos:2018kqj} at NNLL, based on the approach developed in
\refscite{Collins:1981uk,Collins:1981va,Kodaira:1981nh,Kodaira:1982az}.
To compare our formalism to theirs, we start from \eq{EEC_resummed} and choose the resummation scales as
\begin{align} \label{eq:equal_scales}
 \mu_H = \nu_J \equiv \mu_h \sim Q
\,,\qquad
 \mu_J = \mu_S = \nu_S \equiv \mu_l \sim \frac{b_0}{b_T}
\,,\qquad
 \mu_0 = \frac{b_0}{b_T}
\,,\end{align}
i.e.~we distinguish only an overall high scale $\mu_h$ and low scale $\mu_l$,
but always evaluate the rapidity anomalous dimension at its canonical scale $\mu_0 = b_0/b_T$.
With these choices, \eq{EEC_resummed} can be rewritten as
\begin{align} \label{eq:EEC_resummed_simple}
 \frac{\df\sigma}{\df z} &
 = \frac{\born}{8} H_{q\bq}(Q,\mu_h) \int_0^\infty \!\!\df (b_T Q)^2 \, J_0\bigl(b_T Q \sqrt{1-z}\bigr)
   J_q\Bigl(b_T, \mu_l, \frac{\mu_h}{Q}\Bigr) J_\bq\Bigl(b_T, \mu_l, \frac{\mu_h}{Q}\Bigr) \tilde S_q(b_T, \mu_l, \mu_l)
\nn\\&\quad \times
   \exp\left\{ -\int_{\mu_l^2}^{\mu_h^2} \frac{\df\mu'^2}{\mu'^2} \left[ \ln\frac{Q^2}{\mu'^2} A^q[\as(\mu')] + B^q[\as(\mu')] \right] \right\}
\nn\\&\quad \times
   \exp\left\{ \ln\frac{\mu_l^2}{\mu_h^2} \int_{\mu_0^2}^{\mu_l^2} \frac{\df\mu'^2}{\mu'^2}  A^q[\as(\mu')]
             + \ln\frac{Q^2}{\mu_h^2} \frac12 \bigl[ \tilde\gamma_\nu^q[\as(\mu_h)] - \tilde\gamma_\nu^q[\as(\mu_l)] \bigr] \right\}
\,,\end{align}
The coefficients $A^q$ and $B^q$ in \eq{EEC_resummed_simple} are given by
\begin{align} \label{eq:relation_A_B}
 A^i(\as) &= \GammaC^i(\as) + \frac14 \beta(\as) \frac{\df\tilde\gamma_\nu^i(\as)}{\df\as}
\,,\qquad
 B^i(\as) = 2 \gamma_i(\as) - \frac12 \tilde\gamma_\nu^i(\as)
\,,\end{align}
where we remind the reader that $\tilde\gamma_\nu^i[\as(b_0/b_T] \equiv \tilde\gamma_\nu^i(b_T, b_0/b_T)$ is the boundary term of the rapidity anomalous dimension.
Notably, the $A^q$ coefficient differs from the cusp anomalous dimension starting at $\cO(\as^3)$,
which already contributes at NNLL~\cite{Becher:2010tm}.

In \eq{EEC_resummed_simple}, the first line contains the fixed-order boundary terms,
which at canonical scales are free of any logarithms and only depend on $\as(\mu_h) = \as(Q)$ and $\as(\mu_l) = \as(b_0/b_T)$.
The second line in \eq{EEC_resummed_simple} contains the Sudakov form factor that exponentiates the large logarithms.
The third line only contributes when $\mu_h \ne Q$ or $\mu_l \ne \mu_0$, i.e.~when scales are not chosen exactly canonically,
and thus can be used to assess resummation uncertainties by separately varying $\mu_h$ and $\mu_l$.
We note that this procedure is not quite as refined as the one introduced in \sec{EEC_numerics_setup}, where we separately vary \emph{all} resummation scales.
Also note that since $\tilde\gamma_\nu^i = \cO(\as^2)$, the second term in this exponential first contributes at $\cO(\as^3)$.

To compare \eq{EEC_resummed_simple} to the results used \refscite{deFlorian:2004mp,Tulipant:2017ybb,Kardos:2018kqj},
we now explicitly choose the canonical scales, using which \eq{EEC_resummed_simple} reads
\begin{align} \label{eq:EEC_resummed_simple_canonical}
 \frac{\df\sigma}{\df z} &
 = \frac{\born}{8} H_{q\bq}(Q,\mu_h) \int_0^\infty \!\!\df (b_T Q)^2 \, J_0\bigl(b_T Q \sqrt{1-z}\bigr)
   J_q\Bigl(b_T, \mu_l, \frac{\mu_h}{Q}\Bigr) J_\bq\Bigl(b_T, \mu_l, \frac{\mu_h}{Q}\Bigr) \tilde S_q(b_T, \mu_l, \mu_l)
 \nn\\&\qquad \times
 \exp\left\{ -\int_{b_0^2/b_T^2}^{Q^2} \frac{\df\mu'^2}{\mu'^2} \left[ \ln\frac{Q^2}{\mu'^2} A^q[\as(\mu')] + B^q[\as(\mu')] \right] \right\}
\,.\end{align}
For comparison, the resummation formula given in \refcite{Tulipant:2017ybb} reads%
\footnote{Their expansion of the Sudakov form factor contains an implicit $\mu_R$ dependence,
which formally cancels with the $\mu_R$ dependence of the overall hard function.}
\begin{align} \label{eq:EEC_resummed_Grazzini}
 \frac{\df\sigma}{\df z} &
 = \frac{\sigma_{\rm tot}}{8} \tilde H[\as(\mu_R)] \int_0^\infty \df (Q b_T)^2 J_0\bigl(b_T Q \sqrt{1-z}\bigr)
 \nn\\&\qquad\times
 \exp\left\{ -\int_{b_0/b_T^2}^{Q^2} \frac{\df\bar\mu^2}{\bar\mu^2} \left[ \ln\frac{Q^2}{\bar\mu^2} A^q[\as(\bar\mu)] + B^q[\as(\bar\mu)] \right] \right\}
\,,\end{align}
where $\sigma_{\rm tot}$ is the total hadronic cross section.
Comparing \eqs{EEC_resummed_simple}{EEC_resummed_Grazzini}, we first notice that both formulas contain the same Sudakov form factor.%
\footnote{Note that \refcite{deFlorian:2004mp} did not use the correct N$^3$LO result for $A^q$, which was first obtained in \refcite{Becher:2010tm}.}
However, as already discussed at the end of \sec{fact_EEC_q}, their result only contains a combined function $\tilde H$
instead of separating physics at the high and low scales into a hard function and jet and soft functions, respectively.
It obeys
\begin{align} \label{eq:relation_hard}
 \sigma_{\rm tot} \tilde H[\as(Q)] &= \born H_{q\bq}[\as(Q)] J_q[\as(b_0/b_T)] J_\bq[\as(b_0/b_T)]  \tilde S_q[\as(b_0/b_T)]
 \nn\\&\quad
 \times \bigl\{ 1 + \cO[\as(Q)]^2 \bigr\}
\,.\end{align}
Here, each term is evaluated at canonical scales, and thus only depends on the scale through the running coupling.
\Eq{relation_hard} is to be understood as a reexpansion in $\as(b_0/b_T) = \as(Q) + \cO(\as^2)$.
Due to \eq{relation_hard}, both \eqs{EEC_resummed_simple_canonical}{EEC_resummed_Grazzini} recover the correct fixed-order
expansion of the EEC in the back-to-back limit.
However, only \eq{EEC_resummed_simple_canonical} yields the correct NNLL result,
as \eq{EEC_resummed_Grazzini} does not contain the correct boundary terms.

As remarked earlier, while the original works in \refcite{Collins:1981uk,Collins:1981va,Kodaira:1981nh,Kodaira:1982az}
did not yet contain separate hard and jet functions, as they do not yet contribute a the NLL accuracy they work at,
the existence of these functions can already be seen in the $q_T$ factorization
used in those works to derive the EEC factorization in the back-to-back limit.
For instance, \refcite{Kodaira:1982az} also explicitly mentions corrections to the TMDFF in $\as(1/b_T)$.

Another crucial difference lies in the estimation of perturbative uncertainties.
In our approach, we vary all resummation scales, which in particular separately varies the
high scale $\mu_h \sim Q$ and the low scale $\mu_l \sim 1/b_T$. This reflects that in the back-to-back limit,
the EEC contains two parametrically different scales, and varying both resummation scales probes the physics at both scales.
In contrast, in \refscite{deFlorian:2004mp,Tulipant:2017ybb,Kardos:2018kqj} effectively only the hard scale $\mu_R$ is varied,
while the low scale is always kept canonical at $\mu_l = \mu_0 = b_0/b_T$.
This completely neglects the variation from the last line in \eq{EEC_resummed_simple},
and thus largely underestimates the perturbative uncertainties.
Note that due to the lack of a jet and soft function evaluated at the low scale,
variations of the low scale can not even cancel formally in \eq{EEC_resummed_Grazzini},
in contrast to variations of the hard scale $\mu_R$.
Also note that \refscite{deFlorian:2004mp,Tulipant:2017ybb,Kardos:2018kqj} avoid the Landau pole
by deforming the integration contour into the complex plane, rather than freezing out the scale as done in our analysis.

The above observation already explains why the perturbative uncertainties observed in our analysis
are larger than those seen in \refscite{deFlorian:2004mp,Tulipant:2017ybb,Kardos:2018kqj},
as we cover a larger set of scale variations.
Moreover, since variations of the low scale $\mu_l \sim 1/b_T$ probe the strong coupling
at much larger values than variations of the high scale $\mu_h \sim Q$, it is not surprising
that the former are in fact the dominant uncertainties.
To validate this, we compare three methods of estimating uncertainties:
\begin{enumerate}
 \item $\Delta_{\rm tot}$: Full set of profile scale variations as discussed in \sec{EEC_numerics_setup}
 \item $\Delta(\mu_h)$: We only consider two variations,
       \begin{align}
        \mu_H = \nu_J = \frac12 Q \,, \qquad \mu_H = \nu_J = 2 Q
       \,,\end{align}
       while all other scales are kept as in \eq{scales}.
       This roughly mimics the procedure in \refscite{deFlorian:2004mp,Tulipant:2017ybb,Kardos:2018kqj}.
 \item $\Delta(\mu_l)$:  We only consider three variations,
       \begin{align}
         (\mu_J, \mu_S) = \Bigl(\frac12, 1\Bigr) \frac{b_0}{b_T^*} \,,\quad
         (\mu_J, \mu_S) = \Bigl(1, \frac12\Bigr) \frac{b_0}{b_T^*} \,,\quad
         (\mu_J, \mu_S) = \Bigl(\frac12, \frac12\Bigr) \frac{b_0}{b_T^*}
       \,,\end{align}
       while all other scales are kept as in \eq{scales}.
       These are the only scale variations where the hard scales $\mu_{H,J}$ are unchanged,
       and the low scales are only varied down. We also keep the rapidity scale $\nu_S = b_0/b_T^*$ canonic,
       as it does not enter the running coupling.
\end{enumerate}

\begin{figure*}
 \centering
 \includegraphics[width=0.48\textwidth]{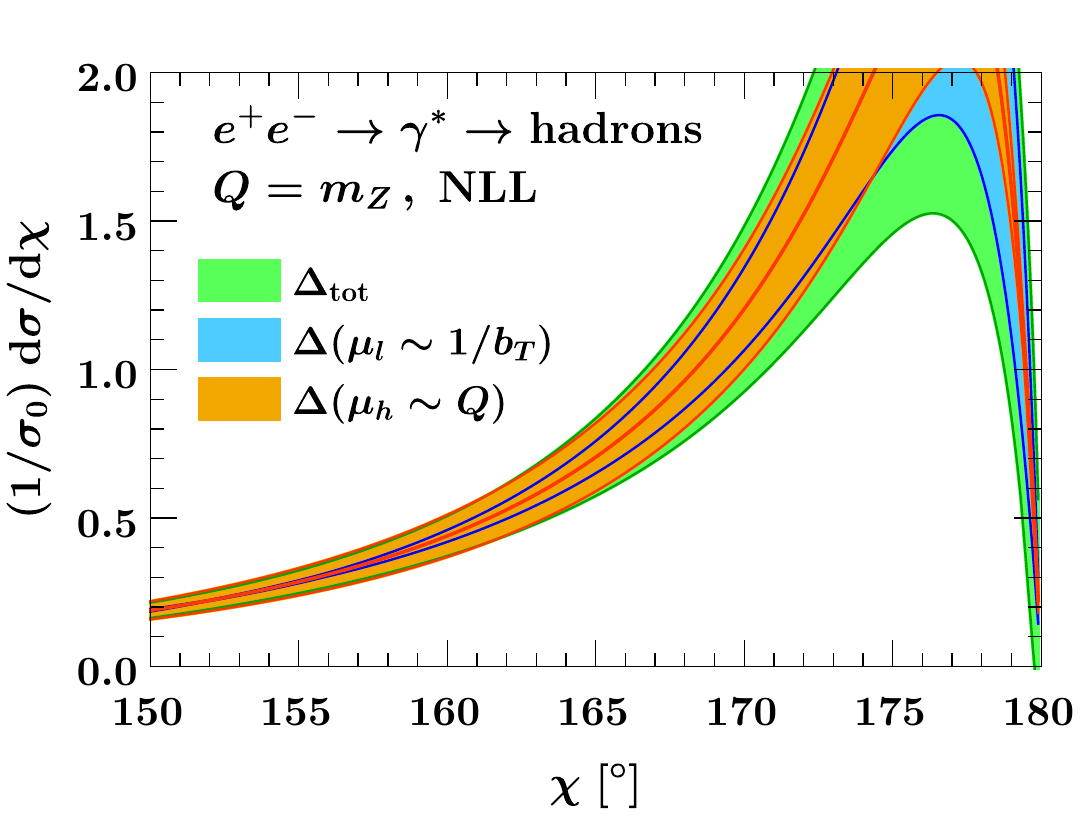}
 \hfill
 \includegraphics[width=0.48\textwidth]{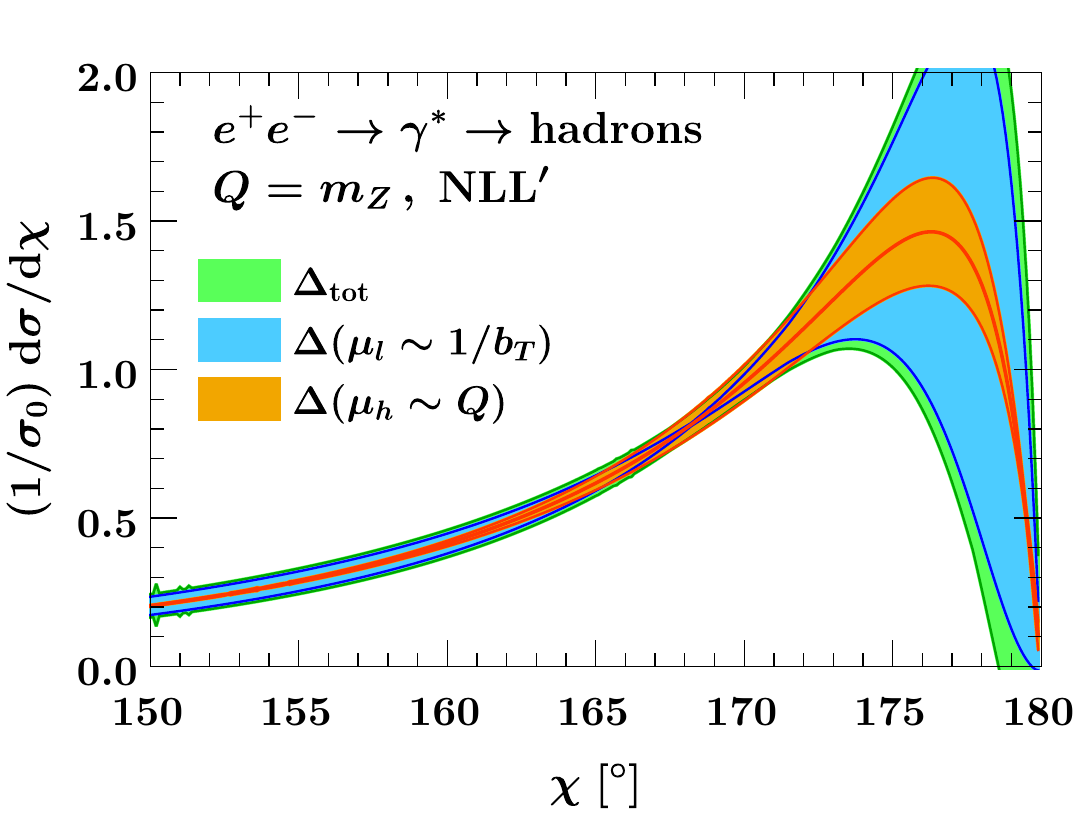}
 \\
 \includegraphics[width=0.48\textwidth]{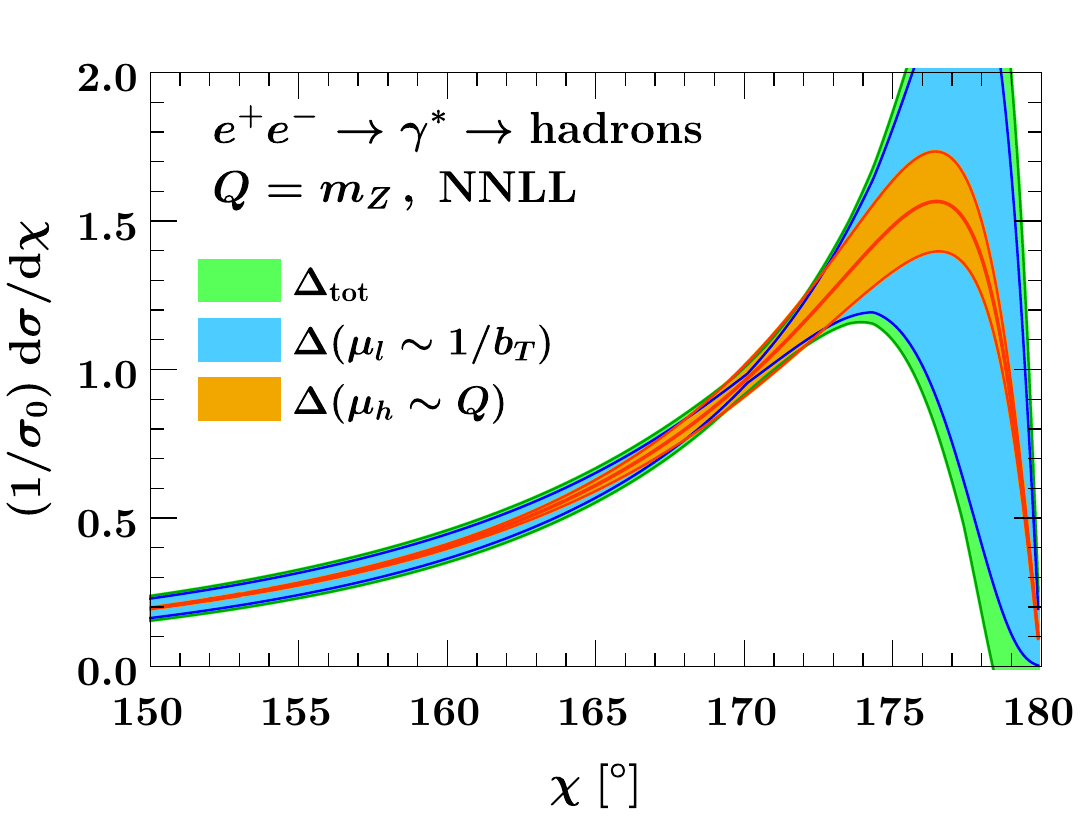}
 \hfill
 \includegraphics[width=0.48\textwidth]{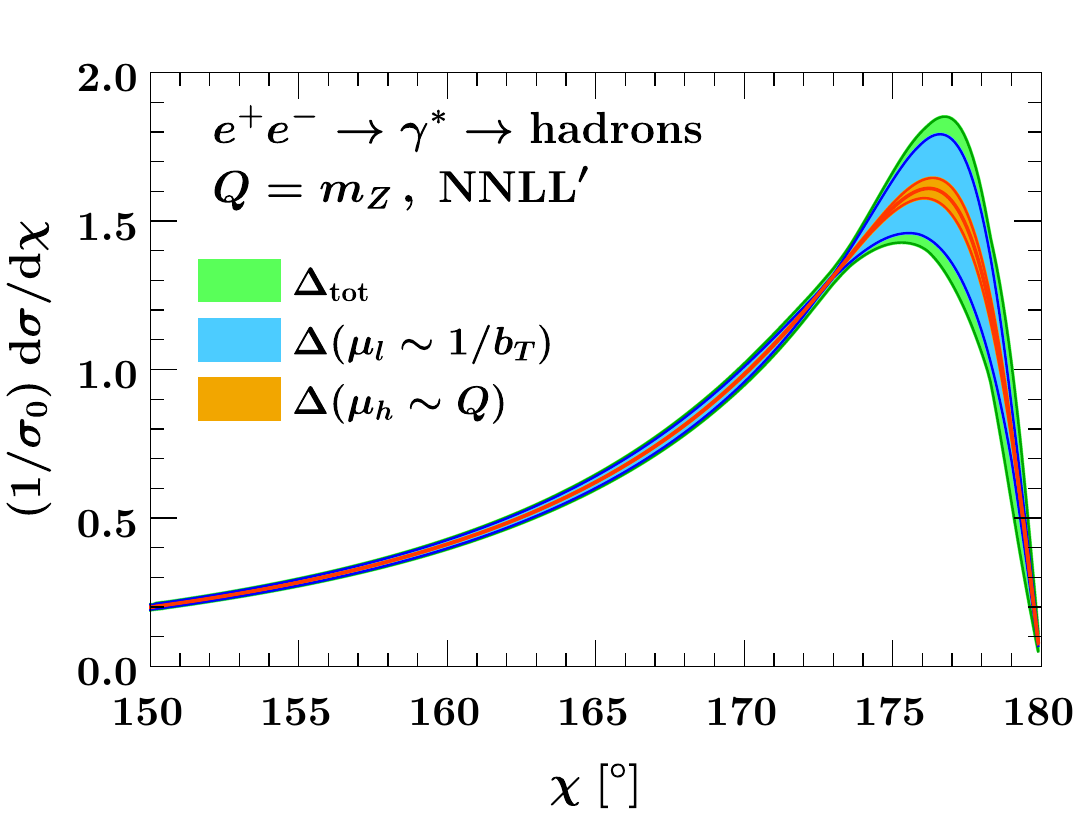}
 \\
 \includegraphics[width=0.48\textwidth]{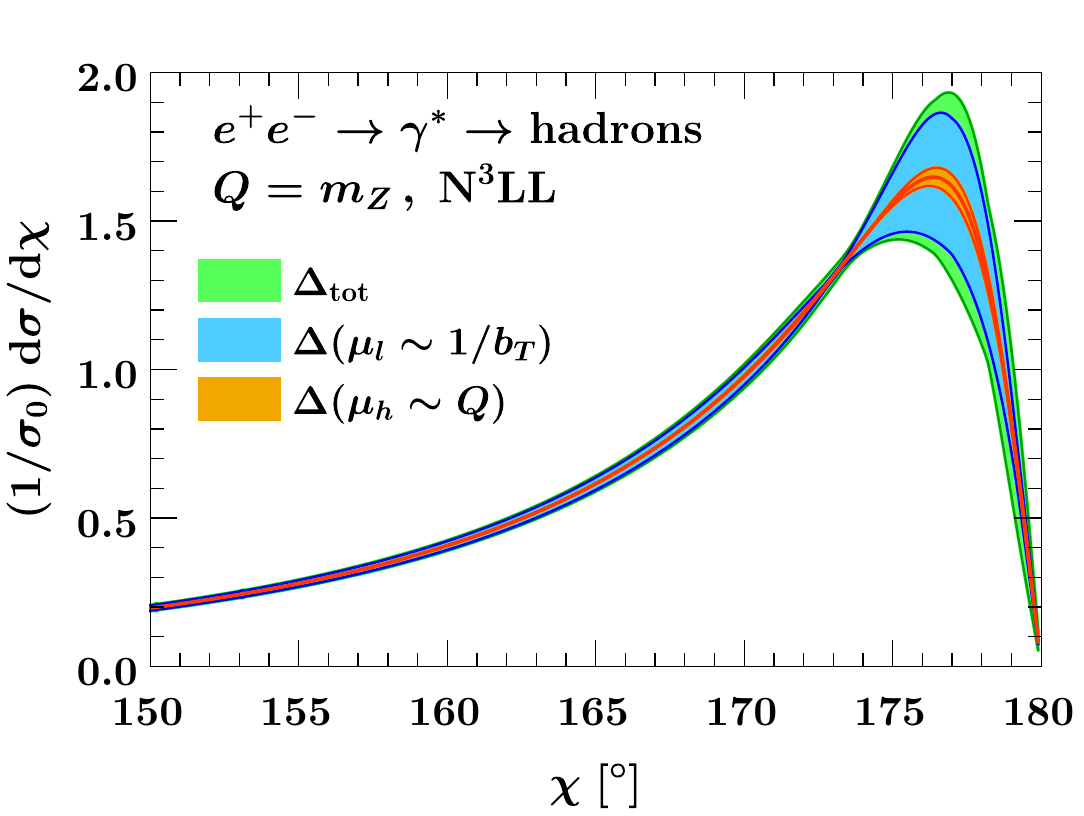}
 \hfill
 \includegraphics[width=0.48\textwidth]{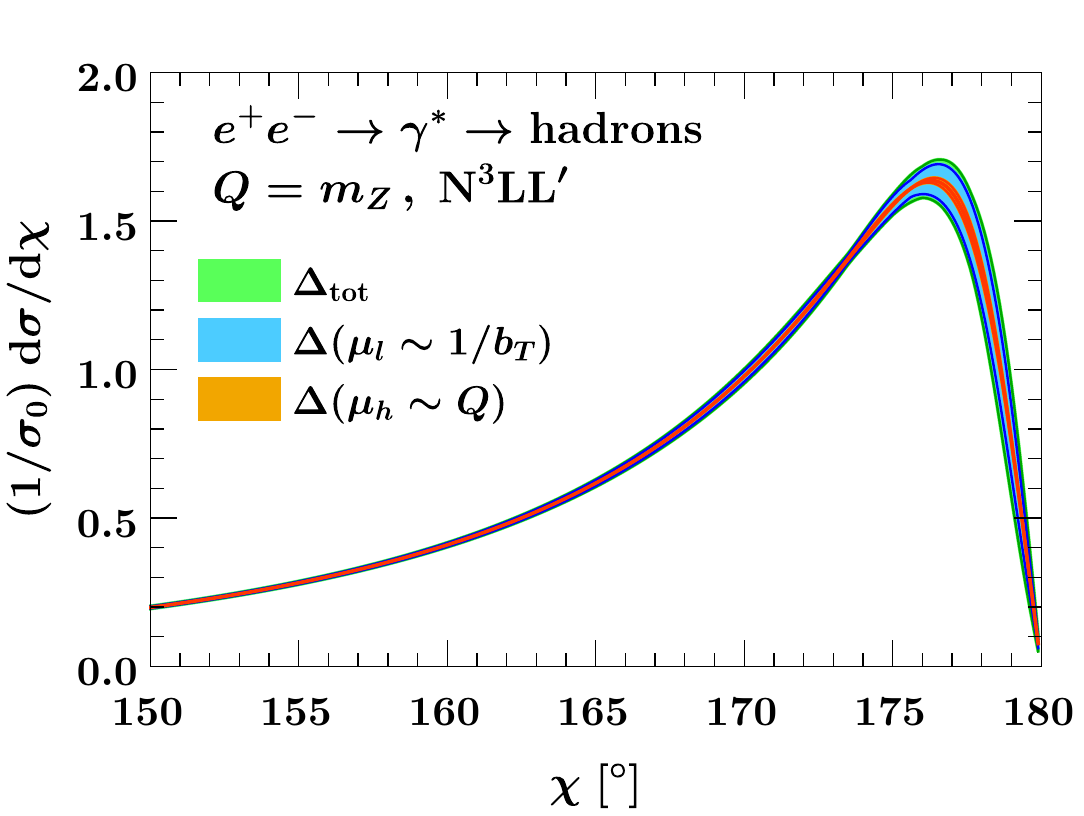}
 \caption{Comparison of the total uncertainty $\Delta_{\rm tot}$, estimated by a full set of profile scale variations,
          to varying only the high scales $\mu_h \sim Q$ or the low scales $\mu_l \sim 1/b_T$.
          See text for more details.}
 \label{fig:EEC_fake_uncertainties}
\end{figure*}

The results are shown in \fig{EEC_fake_uncertainties}, in a similar pattern as in \fig{EEC_uncertainties} such that one can easily compare the two figures.
At NLL, we see that the variations of both the high scale (orange) and low scale (blue) are significant,
giving rise to a very large overall uncertainty (green).
In contrast, at all higher orders we clearly see that it is indeed the low-scale variation $\Delta(\mu_l)$ (blue)
that dominates the total uncertainty, while the high-scale variation $\Delta(\mu_h)$ and the remaining scale variations
are almost negligible.

In \refcite{deFlorian:2004mp}, the uncertainty of the peak at NNLL was given by roughly $\pm 8\%$, compared to $\Delta(\mu_h) \sim 12\%$.
By only considering the variation $\Delta(\mu_h)$, our uncertainty at N$^3$LL$^\prime$ (N$^3$LL) reduces to only $\pm 0.5\%$ ($\pm 2\%$),
compared to our more conservative uncertainty of about $\pm4\%$ ($\pm15\%$) when using all scale variations.
In both approaches, the uncertainty reduces by a factor of about $4$ when going from N$^3$LL to N$^3$LL$^\prime$,
illustrating the importance of including the N$^3$LO boundary terms computed in this work.
However, we stress again that only varying the high scales neglects important uncertainties from soft physics,
and thus the $\Delta(\mu_h)$ variation alone is not sufficient to obtain a robust estimate of theory uncertainties.

We close by remarking that it was already remarked in \refcite{Kodaira:1982az} that the EEC
is quite sensitive to the high $b_T$ region, and nonperturbative model functions were introduced
in order to achieve agreement with CELLO data. This is consistent with our observation
that variations of the $1/b_T$ scales yield the dominant uncertainties.

\FloatBarrier
\section{Conclusions}
\label{sec:conclusion}

In this work we have calculated the full singular structure of the Energy-Energy Correlation (EEC) in the back-to-back limit at $\cO(\alpha_s^3)$ in QCD, including contact terms.
Our work applies both in the case of $e^+e^-$ annihilation as well as in gluon induced Higgs decays.
To obtain these results we have calculated the quark and gluon jet functions for the EEC in the back-to-back limit at N$^3$LO, which were the last missing ingredients for the factorization theorem in this limit at N$^3$LO. 

The computation of the jet functions relies on the calculation of the kernels of the transverse-momentum dependent fragmentation functions at N$^3$LO in our companion paper~\cite{Ebert:2020qef} 
 which have been obtained using a recently developed method for the expansion of cross sections around the collinear limit \cite{Ebert:2020lxs}.
We checked that the logarithmically enhanced terms of our calculation match those predicted by the rapidity renormalization group (RRG) evolution.

By comparing the leading transcendental part of our results we show that the principle of maximal transcendentality, which states that a quantity obtained in $\cN=4$ SYM constitutes the leading transcendental term of the same quantity in QCD, holds for the EEC in the back-to-back asymptotic up to $\cO(\alpha_s^3)$. 
In particular, we show that the leading transcendental part of the EEC in $e^+e^-$ is identical to the one for the EEC in Higgs decay to gluons and that both match the result in $\cN=4$ \cite{Henn:2019gkr,Korchemsky:2019nzm,Kologlu:2019mfz}, not only for the logarithmic part but also for the contact terms. This also provides a non-trivial cross check on both the quark and gluon jet function calculations of the $\delta(1-z)$ constants in addition to the one on the logarithmic parts coming from the RRG evolution.

Leveraging on the fact that the EEC obeys a set of non trivial sum rules~\cite{Korchemsky:2019nzm,Kologlu:2019mfz}, we used our newly calculated result for the contact terms at $\cO(\alpha_s^3)$ as well as the logarithmic enhanced contributions in the small angle limit \cite{Dixon:2019uzg} to obtain the \moment of the EEC distribution in the bulk at NNLO in QCD analytically. This constitutes the first piece of analytic information on the EEC distribution at this order in QCD away from the endpoints, both in the case of $e^+e^-$ annihilation as well as in gluon induced Higgs decays.

Finally, we have carried out the resummation of the EEC in the back-to-back region at N$^3$LL$^\prime$ accuracy.
This is the first time an event shape observable is resummed at this level of accuracy and, more generally, this constitutes the highest level of resummation for any infrared and collinear safe observable in QCD to date.
We show that by performing the resummation at N$^3$LL$^\prime$ we obtain a reduction of uncertainties by a factor of $\sim 4$ in the peak region compared to previous results obtained at lower accuracy.
We thoroughly discuss different schemes to estimate the uncertainties due to missing higher order corrections both in the boundary terms as well as in the anomalous dimensions.
We compare these different schemes with the ones used in the literature for this observable. Adopting a scheme in line with the ones previously used in the literature we obtain a 5 per-mille uncertainty at the peak. Using a more conservative scheme, which includes a significant contribution from non-perturbative regions, we obtain a 4\% uncertainty for our result at N$^3$LL$^\prime$.
We point out that the accuracy for lower order results is severely affected by the choice of scheme and that varying the low-energy scale gives a dramatically larger estimate of uncertainties.

The recent progress in understanding Energy-Energy correlators is very promising and we believe it shows that they will play a crucial role in improving our understanding of the strong interaction in the years to come.
For example, as the EEC has been often used to determine the strong coupling constant \cite{Abreu:1990us,Acton:1991cu,Abreu:1993kj,Abe:1994mf,Tulipant:2017ybb,Kardos:2018kqj,dEnterria:2019its}, it would be interesting to leverage the high level of perturbative control we gained on this observable thanks to the N$^3$LL$^\prime$ resummation, to improve the extraction of $\alpha_s$, complementing the extractions based on other event shape observables in $e^+ e^-$ such as thrust and C-parameter \cite{Becher:2008cf,Abbate:2010xh,Bethke:2011tr,Hoang:2015hka}.
In addition, it could also be used to extract nonperturbative corrections to the rapidity anomalous dimensions.
We expect that further studies of perturbative~\cite{Moult:2019vou} and non-perturbative~\cite{Korchemsky:1999kt,Li:2021txc} power corrections will be important to improve the theoretical control on this observable.

It will also be interesting to explore the application of the techniques used in this work and its companion paper~\cite{Ebert:2020qef} to higher point energy correlators in QCD where recent progress has been obtained \cite{Chen:2019bpb,Chen:2020vvp,Chen:2020adz}.

\acknowledgments
We thank Ian Moult, Lance Dixon and HuaXing Zhu for useful discussions on the EEC, and G\'abor Somogyi for correspondance on the numerical results of \refcite{Tulipant:2017ybb}.
This work was supported by the Office of High Energy Physics of the U.S. DOE under Contract No.\ DE-AC02-76SF00515 and by the Office of Nuclear Physics of the U.S.\ DOE under Contract No.\ DE-SC0011090, and within the framework of the TMD Topical Collaboration.
M.E.\ is also supported by the Alexander von Humboldt Foundation through a Feodor Lynen Research Fellowship,
and B.M.\ is also supported by a Pappalardo fellowship.

\appendix

\section{Fixed-order structure of the EEC jet function}
\label{app:EEC_jet_fixed_order}

Here, we provide the fixed-order structure of the jet function, obtained by solving the RG \eqs{jet_RGEs}{jet_anom_dims} order-by-order in $\as$.
The fixed-order coefficients as defined by \eq{J_expansion} are given by
\begin{align} \label{eq:J_coeffs}
 J_i^{(0)}(L_b, L_Q) &
 = 1
\,,\nn\\
 J_i^{(1)}(L_b, L_Q) &
 = L_b \Bigl(\Gamma^i_0 L_Q + \frac{\tilde\gamma^i_{J\,0}}{2}\Bigr) + j_i^{(1)}
\,,\nn\\
 J_i^{(2)}(L_b, L_Q) &
 = \frac12 L_b^2 \Bigl[ L_Q^2 (\Gamma^i_0)^2 + \Gamma^i_0 L_Q  \Bigl(\tilde\gamma^i_{J\,0}+\beta_0\Bigr) + \frac{1}{4} \tilde\gamma^i_{J\,0} \Bigl(\tilde\gamma^i_{J\,0}+2 \beta_0\Bigr) \Bigr]
   \nn\\&\quad
 + L_b \Bigl[ L_Q \bigl(\Gamma^i_1+ \Gamma^i_0 j_i^{(1)} \bigr) + \frac{1}{2} \tilde\gamma^i_{J\,1} + \Bigl( \frac{1}{2} \tilde\gamma^i_{J\,0} + \beta_0 \Bigr) j_i^{(1)} \Bigr]
   \nn\\&\quad
 - \frac{\tilde\gamma^i_{\nu\,1}}{2} L_Q + j_i^{(2)}
\,,\nn\\
 J_i^{(3)}(L_b, L_Q) &
 = \frac16 L_b^3 \Bigl[ L_Q^3 (\Gamma^i_0)^3
   + \frac32 (\Gamma^i_0)^2 L_Q^2  (\tilde\gamma^i_{J\,0}+2 \beta_0)
   + \Gamma^i_0 L_Q  \Bigl( \frac34 (\tilde\gamma^i_{J\,0})^2 + 3 \beta_0 \tilde\gamma^i_{J\,0} + 2 \beta_0^2 \Bigr)
   \nn\\&\qquad\quad
   + \tilde\gamma^i_{J\,0} \Bigl( \frac34  \beta_0 \tilde\gamma^i_{J\,0} + \frac18 (\tilde\gamma^i_{J\,0})^2 + \beta_0^2 \Bigr) \Bigr]
 \nn\\&\quad
  + L_b^2 \Bigl\{
   L_Q^2 \Gamma^i_0 \Bigl(\Gamma^i_1 + \frac12 \Gamma^i_0 j_i^{(1)} \Bigr)
   \nn\\&\qquad\qquad
   + \frac{1}{2} L_Q \Bigl[  \Gamma^i_1 (\tilde\gamma^i_{J\,0} + 2 \beta_0) + \Gamma^i_0 (\beta_1 + \tilde\gamma^i_{J\,1}) +\Gamma^i_0 (\tilde\gamma^i_{J\,0} + 3 \beta_0) j_i^{(1)} \Bigr]
   \nn\\&\qquad\qquad
   + \frac12 \beta_0 \tilde\gamma^i_{J\,1} +  \frac14 \tilde\gamma^i_{J\,0} (\tilde\gamma^i_{J\,1} + \beta_1)  + \Bigl(\frac34 \beta_0 \tilde\gamma^i_{J\,0} + \frac18 (\tilde\gamma^i_{J\,0})^2 + \beta_0^2 \Bigr) j_i^{(1)}
 \Bigr\}
 \nn\\&\quad
  + L_b \Bigl\{ - \frac{1}{2} \Gamma^i_0 \tilde\gamma^i_{\nu\,1} L_Q^2
   + L_Q \Bigl[ \Gamma^i_2   + \Gamma^i_1 j_i^{(1)} + \Gamma^i_0 j_i^{(2)} - \Bigl(\frac14 \tilde\gamma^i_{J\,0} + \beta_0 \Bigr) \tilde\gamma^i_{\nu\,1} \Bigr]
   \nn\\&\qquad\qquad
   + \frac12\tilde\gamma^i_{J\,2}  + \Bigl(\frac12 \tilde\gamma^i_{J\,1} + \beta_1\Bigr) j_i^{(1)} + \Bigl(\frac12 \tilde\gamma^i_{J\,0} + 2 \beta_0 \Bigr) j_i^{(2)}
   \Bigr\}
 \nn\\&\quad
  - \frac{1}{2} L_Q \bigl( \tilde\gamma^i_{\nu\,2} +  \tilde\gamma^i_{\nu\,1} j_i^{(1)} \bigr) + j_i^{(3)}
\,.\end{align}
Here, the $\Gamma_n^i$ and $\gamma_{J\,n}^i$ are the $\cO[(\as/4\pi)^n]$ coefficients of the cusp and jet noncusp anomalous dimensions, respectively,
where we remind the reader that the jet noncusp anomalous dimension is identical to that of the TMD beam function,
$\gamma_{J\,n}^i \equiv \gamma_{B\,n}^i$.
Explicit expressions for these anomalous dimensions in our conventions are collected in~\refcite{Billis:2019vxg}.
The corresponding fixed-order expansion of the polarized gluon jet function $J'_g$
can be obtained from \eq{J_coeffs} by dropping all terms that do not contain an explicit factor $j_i^{(n)}$,
as in this case $J_g^{\prime(0)} = 0$.

\section{Bessel transform}
\label{app:EEC_Bessel}

To evaluate \eqs{EEC_fact_thm_q}{EEC_fact_thm_g} at fixed order, we need to evaluate Bessel transforms of the form
\begin{align} \label{eq:FT_1}
 I_n &\equiv
 \frac18 \int\df(b_T^2 Q^2) J_0\bigl(b_T Q \sqrt{\zb}\bigr) \, \ln^n\frac{b_T^2 \mu^2}{b_0^2}
 \nn\\&
 = \frac12 \sum_{k=0}^{n-1} (-1)^{k+1} n \binom{n-1}{k} R_2^{(n-k-1)} \biggl[ \frac{\ln^k[(Q^2/\mu^2)\zb]}{\zb} \biggr]_+ + \frac12 R_2^{(n)} \delta(\zb)
\,,\end{align}
which follows immediately from Eq.~(C.16) of \refcite{Ebert:2016gcn}, with
\begin{align}
 R_2^{(n)} = \frac{\df^n}{\df a^n} e^{2\gamma_E a} \frac{\Gamma(1+a)}{\Gamma(1-a)} \biggr|_{a=0}
\,.\end{align}
The plus distributions in \eq{FT_1} are defined as usual such that $\int_0^1 \df x \, [\ln^n x/x]_+ = 0$.
They can be easily rewritten in terms of distributions in $\zb$ using
\begin{align}
 \biggl[ \frac{\ln^k[(Q^2/\mu^2)\zb]}{\zb} \biggr]_+
 &= \sum_{\ell=0}^k \binom{k}{\ell} \ln^\ell\frac{Q^2}{\mu^2} \cL_{k-\ell}(\zb)
 + \frac{\ln^{k+1}(Q^2/\mu^2)}{k+1} \delta(\zb)
\,,\end{align}
where the plus prescription on the right hand side now acts with respect to $\zb$.
For example, the first few Bessel transforms read
\begin{align} \label{eq:FT_N3LO}
 I_0 &= \frac{1}{2} \delta(\zb)
\,,\nn\\
 I_1 &= -\frac{1}{2} \bigl[\cL_0(\zb) + L_h \delta(\zb) \bigr]
\,,\nn\\
 I_2 &= \cL_1(\zb) + L_h \cL_0(\zb) + \frac12 L_h^2 \delta(\zb)
\,,\nn\\
 I_3 &= -\frac32 \cL_2(\zb) - 3 L_h \cL_1(\zb) - \frac32 L_h^2 \cL_0(\zb) + \Bigl( \frac12 L_h^3 - 2 \zeta_3\Bigr) \delta(\zb)
\,,\end{align}
where $L_h = \ln(Q^2/\mu^2)$.
Starting from $n=3$, the $R_2^{(n)}$ terms start to induce $\zeta$ values.

\section{Sum Rules Ingredients}\label{app:sumruleIngredients}

Here we collect all required results for the quantities in \eq{constDef}.

First, we note that $V_1^{(n)}$ is the coefficient of $\delta(\zb)$ at $\cO(\alpha_s^n)$,
and thus can be immediately read off from the results in \eqs{EEC_z1_NLO_NNLO}{EEC_z1_N3LO_ee} for $e^+e^-$
and in \eqs{EEC_z1_NNLO_H}{EEC_z1_N3LO_H} for Higgs.
For example,
\beq
	V_{1,e^+e^-}^{(1)} = - C_F (4 + 2 \zeta_2) \,,\qquad V_{1,H}^{(1)}=C_A \left(\frac{65}{18}-2  \zeta_2\right)-\frac{5}{18}n_f\,.
\eeq

The expressions for $V_0^{(n)}$, the coefficients of $\delta(z)$, can be in principle extracted via the factorization formula for the EEC in the collinear limit of \refcite{Dixon:2019uzg}. Here, we have extracted them by using the sum rule in \eq{sumRuleNorderN} and analytically integrating the regular part of the distribution obtained from the fixed order calculation of \refscite{Dixon:2018qgp,Luo:2019nig}.
For $e^+e^-$ we obtain
\begin{align}
V_{0}^{(1)}&=\frac{13}{24} C_F\,,
\nn\\
V_{0}^{(2)}&=
C_F^2 \left(44 \zeta_4-84 \zeta_3+\frac{109}{12}\zeta_2+\frac{208219}{5184}\right)+C_F n_f \left(\frac{33 \zeta_2}{20}+4 \zeta_3-\frac{751777}{216000}\right)
\nn\\&+ 
C_F C_A \left(-22 \zeta_4+36 \zeta_3-\frac{373}{45}\zeta_2-\frac{3032011}{162000}\right)\,,
\end{align}
while for the Higgs case we obtain:
\begin{align}
V_{0}^{(1)}&=\frac{3311}{300}C_A-\frac{1447}{600}n_f\,,
\nn\\
V_{0}^{(2)}&=
C_A^2 \left(22 \zeta_4-\frac{242}{3}\zeta_3-\frac{18337}{450}\zeta_2+\frac{31760837}{101250}\right)-C_A n_f \left(\frac{4}{15}\zeta_3-\frac{2891}{225}\zeta_2+\frac{183628817}{1620000}\right)
\nn\\&+
	C_F n_f \left(\frac{224}{15}\zeta_3-\frac{23}{10}\zeta_2-\frac{249491}{12000}\right)-n_f^2
   \left(\frac{22}{15}\zeta_2-\frac{274091}{27000}\right)
\end{align}

Finally, for the expressions for the total cross section we take the results of \refcite{Herzog:2017dtz}. For $e^+e^-$ they read%
\footnote{Note that for the color structure $d_{abc}d^{abc}$ of the singlet part we adopted the same convention as in \eq{EEC_z1_N3LO_ee}, which is different from the one adopted in \refcite{Herzog:2017dtz}.}
\begin{align}
R_{e^+e^-}^{(1)}&= 3 C_F
\nn \\
R_{e^+e^-}^{(2)}&= - { 3 \over 2 }\, \* C_F^2 \,+\, C_A \* C_F \, \*  \Bigg({ 123 \over 2 }  - 44\, \* \zeta_3  \Bigg) \,-\, C_F \* n_f \, \*  \Big(11- 8\, \* \zeta_3\Big)
\nn \\
R_{e^+e^-}^{(3)}&=   - { 69 \over 2 }\, \* C_F^3 
       \,- \,C_A \* C_F^2 \, \*  \Big(
            127
          + 572\, \* \zeta_3
          - 880\, \* \zeta_5
          \Big)
\nn \\ &
       + \,C_A^2 \* C_F \, \*  \Bigg(
            { 90445 \over 54 }
          - { 242 \over 3 }\, \* \zeta_2
          - { 10948 \over 9 }\, \* \zeta_3
          - { 440 \over 3 }\, \* \zeta_5
          \Bigg)
       \,-\,C_F^2\, \* n_f \, \*  \Bigg(
            { 29 \over 2 }
          - 152\, \* \zeta_3
          + 160\, \* \zeta_5
          \Bigg)
\nn \\ &
       - \,C_A \* C_F\, \* n_f \, \*  \Bigg(
            { 15520 \over 27 }
          - { 88 \over 3 }\, \* \zeta_2
          - { 3584 \over 9 }\, \* \zeta_3
          - { 80 \over 3 }\, \* \zeta_5
          \Bigg)
       \,+\,C_F\, \* n_f^2 \, \*  \Bigg(
            { 1208 \over 27 }
          - { 8 \over 3 }\, \* \zeta_2
          - { 304 \over 9 }\, \* \zeta_3
          \Bigg)\,.
\nn \\ &
+ N_{F,V} \frac{d_{abc}d^{abc}}{16N_r} \left(\frac{176}{3} - 128\zeta_3\right)\,.
\end{align}
For Higgs, they are given by
\begin{align}
R_H^{(1)} &={73 \over 3}\, \* C_A
   - {14 \over 3}\, \* n_f
\nn \\
R_H^{(2)} &=
       C_A^2 \: \* \Bigg(
          {37631 \over 54}
        - {242 \over 3}\, \* \zeta_2
        - 110\, \* \zeta_3
        \Bigg)
     \: - \: C_A \* \,n_f \: \* \Bigg(
          {6665 \over 27}
        - {88 \over 3}\, \* \zeta_2
        + 4\, \* \zeta_3
        \Bigg)
\nn \\ &
     - \: C_F \* \,n_f \, \* \Bigg(
          {131 \over 3}
        - 24\, \* \zeta_3
        \Bigg)
     \: + \: n_f^2 \: \* \Bigg(
          {508 \over 27}
        - {8 \over 3}\, \* \zeta_2
        \Bigg)
\nn \\
R_H^{(3)} &=
         C_A^3 \, \* \Bigg( \,
            {15420961 \over 729}
          - {45056 \over 9}\,\* \zeta_2
          - {178156 \over 27}\,\* \zeta_3
          + {3080 \over 3}\, \* \zeta_5
          \Bigg)
\nn \\ &
       - \: C_A^2 \* \,n_f \, \* \Bigg( \,
            {2670508 \over 243}
          - {8084 \over 3}\, \* \zeta_2
          - {9772 \over 9}\, \* \zeta_3
          + {80 \over 3}\, \* \zeta_5
          \Bigg)
\nn \\ &
       - \: C_F \* C_A \* \,n_f\, \* \Bigg( \,
            {23221 \over 9}
          - {572 \over 3}\, \* \zeta_2
          - 1364\, \* \zeta_3
          - 160\, \* \zeta_5
          \Bigg)
\nn \\ &
       + \: C_F^2 \* \,n_f \, \* \Bigg( \,
            {221 \over 3}
          + 192\, \* \zeta_3
          - 320\, \* \zeta_5
          \Bigg)
       \: + \: C_A \* \,n_f^2 \, \* \Bigg( \,
            {413308 \over 243}
          - {1384 \over 3}\, \* \zeta_2
          + {56 \over 9}\, \* \zeta_3
          \Bigg)
\qquad \nn \\ &
       + \: C_F \* \,n_f^2 \, \* \Big( 
            440\:
          - {104 \over 3}\, \* \zeta_2
          - 240\, \* \zeta_3
          \Big)
       \: - \: n_f^3 \, \* \Bigg( \,
            {57016 \over 729}
          - {224 \over 9}\, \* \zeta_2
          - {64 \over 27}\, \* \zeta_3
          \Bigg)\,.
\end{align}

\addcontentsline{toc}{section}{References}
\bibliographystyle{jhep}
\bibliography{../refs}

\end{document}